\documentclass{emulateapj}
\pdfoutput=1
\usepackage{graphics}
\usepackage{rotating}
\usepackage{ifpdf}

\newcommand{\beq}{\begin{equation}}
\newcommand{\eeq}{\end{equation}}

\newcommand{\teff}{T_{\rm eff}}

\def\earth{\hbox{$\oplus$}}
\def\arcmin{\hbox{$^\prime$}}
\def\arcsec{\hbox{$^{\prime\prime}$}}

\newcommand{\lsim}{\ \raise
-2.truept\hbox{\rlap{\hbox{$\sim$}}\raise5.truept\hbox{$<$}\ }}
\newcommand{\gsim}{\ \raise
-2.truept\hbox{\rlap{\hbox{$\sim$}}\raise5.truept\hbox{$>$}\ }}
\newcommand{\simsim}{\ \raise
-2.truept\hbox{\rlap{\hbox{$\sim$}}\raise5.truept\hbox{$\sim$}\ }}

\slugcomment{Accepted for publication by the Astrophysical Journal}
\shorttitle{VLMs and BDs in Orion}
\shortauthors{Da Rio et al. 2011}

\begin{document}

\title{The Initial Mass Function of the Orion Nebula Cluster across the H-burning limit}

\author{N. Da Rio}
\affil{European Space Agency, Keplerlaan 1, 2200 AG Noordwijk, The Netherlands}
\email{ndario@rssd.esa.int}

\author{M. Robberto}
\affil{Space Telescope Science Institute, 3700 San Martin Dr., Baltimore MD, 21218, USA\ }

\author{L. A. Hillenbrand}
\affil{California Institute of Technology, 1200 East California Boulervard, 91125 Pasadena, CA, USA}

\author{T. Henning}
\affil{Max-Planck-Institut f\"{u}r Astronomie, K\"{o}nigstuhl 17, 69117 Heidelberg, Germany}

\author{K. G. Stassun\altaffilmark{2,3} }
\affil{Vanderbilt Univ., Dept. of Physics \& Astronomy 6301 Stevenson Center Ln., Nashville, TN 37235, USA }
\altaffiltext{2}{Fisk University, Department of Physics, 1000 17th Ave. N., Nashville, TN 37208, USA}
\altaffiltext{3}{Massachusetts Institute of Technology, Department of Physics, 77 Massachusetts Ave., Cambridge, MA 02139, USA}


\begin{abstract}
We present a new census of the Orion Nebula Cluster (ONC)
over a large field of view ($\gtrsim30^\prime\times30^\prime$),
significantly increasing the known population of stellar and substellar
cluster members with precisely determined properties.
We develop and exploit a technique
to determine stellar effective temperatures from optical colors,
nearly doubling the previously available number of objects with
effective temperature determinations in this benchmark cluster.
Our technique utilizes colors from deep photometry in the $I$-band and in
two medium-band filters at $\lambda\sim753$ and $770$~nm, which accurately
measure the depth of a molecular feature present in the spectra of cool
stars. From these colors we can derive effective temperatures with a
precision corresponding to better than one-half spectral subtype,
and importantly this precision is independent of the extinction to the
individual stars. Also, because this technique utilizes only photometry
redward of 750~nm,
the results are only mildly
sensitive to optical veiling produced by accretion.
Completing our census with previously available data, we place some
1750 sources in the Hertzsprung-Russel diagram and assign masses
and ages down to 0.02 solar masses.  At faint luminosities,
we detect a large population of background sources
which is easily separated in our photometry from the bona fide
cluster members.
The resulting initial mass function of the cluster has good completeness
well into the substellar mass range, and we find that it declines steeply
with decreasing mass.
This suggests a deficiency of newly formed brown dwarfs in the cluster
compared to the Galactic disk population.
\end{abstract}

\keywords{stars: formation, pre-main sequence, mass function, brown dwarfs, stellar clusters: individual (Orion Nebula Cluster)  }


\section{Introduction}
\label{section:introduction}
Understanding the initial mass function (IMF) is one of the most important
problems of stellar astrophysics. The IMF, together with the star formation
history, dictates origin, evolution, and fate of the stellar populations,
from individual clusters up to entire galaxies. The distribution of stellar
masses has been studied in depth for more than half a century, starting
with the pioneering work of \citet{salpeter55}; the major open question
is whether the IMF is universal or if it depends
on the initial conditions of star formation \citep[see,
e.g.,][]{kroupascience,bastian2010}. Whereas it is commonly accepted that
the IMF seems well reproduced by a power
law for masses greater than several tenths of a solar mass,
the shape and universality of the IMF in the substellar mass regime
is still under investigation \citep[e.g.,][]{Wang2011}.

Very young clusters (few Myr old) in star-forming regions provide a unique
tool to investigate the IMF across the entire mass spectrum, for a number
of reasons. First, they are usually young enough that neither dynamical
processes nor stellar evolution have altered the mass distribution;
therefore the measured distribution of stellar masses coincides with the
IMF. Moreover, young low-mass stars and brown dwarfs (BDs) are in their brightest
evolutionary stage when they contract towards the main sequence, therefore
these objects are more easily detected and characterized when they are young.

Among the nearby Galactic star-forming regions, the Orion Nebula
Cluster (ONC) is an ideal site for the study of star formation
in particular for low-mass stars and BDs. This cluster counts a
few thousand members, 1-3~Myr old \citep{hillenbrand97,paperII},
spanning the entire mass spectrum ($M\lesssim50$~M$_\odot$). Several
studies have been conducted in the past decade to measure the IMF in the ONC
\citep[e.g.,][]{hillenbrand97,hillenbrand-carpenter2000,muench2000,muench2002,lucasroche2000,lucas2005,slesnick04,paperII};
they generally find (similar to the field star population) a Salpeter-like
slope above 1~M$_\odot$, which flattens to a broad peak at 0.2-0.3~M$_\odot$
(though the shape and position of the IMF peak is highly model dependent,
\citealt{paperII}), and the mass distribution likely decreases in the
substellar mass range.

Determining masses in a young region such as the ONC
is not without difficulties: besides the strong nebular emission,
another major impediment is caused by
differential reddening, which has strongly limited the ability to derive the
stellar parameters of individual sources based on photometry alone. To
overcome this shortcoming, spectroscopic surveys have been carried out
\citep{hillenbrand97,lucas2001,lucas2006,slesnick04,riddick2007,weights2009},
but they are either limited to a fraction of the members, or to the very
central part of the region, the Trapezium cluster. Using near-infrared
(NIR) photometry it is possible to asses stellar masses down to the
planetary masses ($M<13$~M$_J$), for example by de-reddening the measured
color-magnitude diagrams (CMDs) on one isochrone or even to deriving directly
the IMF from the NIR luminosity functions (LFs)
\citep{hillenbrand-carpenter2000,muench2002}. This
second approach, however, is not very accurate: the ONC shows a
significant luminosity spread, sometimes interpreted as a real age
spread \citep{reggiani2011}, and stars of different age follow different
mass-luminosity relation. Moreover the NIR excess originating in the inner
circumstellar disks \citep{meyer97} alters the observed fluxes.

Clearly, in order to improve our knowledge of the ONC IMF, a precise and
systematic characterization of the stellar parameters of individual
members is needed. To this purpose, the \emph{Orion HST Treasury Program}
\citep{robberto2005treasury} has produced a high spatial resolution
photometric survey of the ONC with three instruments onboard the Hubble Space
Telescope, from $U-$band to the NIR, over a large field of view (FOV),
with a sensitivity well into the BD mass range. To fully exploit this exceptional
dataset, an accurate estimate of the effective temperature ($\teff$)
is needed to derive $A_V$ from the observed colors and therefore the
stellar luminosities.  In our previous works \citep[][hereafter Paper~I
and Paper~II]{paperI,paperII} we have collected spectral types from
the literature
and complemented
them with new optical spectroscopy. Moreover, we have presented a new
observational technique, based on optical medium-band photometry at
6200\AA---a wavelength where the spectra of cool stars show a deep,
$\teff$-dependent TiO absorption feature---to derive the spectral types
of M-type sources. This allowed us to obtain the stellar parameters of
$\sim1000$ members and derive the most complete HRD of the ONC, down to
the hydrogen burning limit.

In the present paper we extend our investigation to lower masses, well
into the BD regime, and with higher completeness. To this purpose, we
take advantage of the same observational strategy presented in Paper~I,
using medium-band photometry to derive spectral types of cool sources,
from which $\teff$ and $A_V$ are derived, and place individual sources
in the HRD. Here, instead of the $6200\AA$ filter used in Paper~I,
we select two bands at longer optical wavelengths, in order to increase
our sensitivity for cooler sources such as BDs and further reduce the
influence of extinction.

In Section \ref{section:the_data} we present our new observations,
the data reduction and calibration. In Section \ref{section:analysis}
we define two spectro-photometric indices based on our medium-band
photometry. Using available stellar parameters for a small sample of ONC
very-low mass stars and BDs, we define empirical transformations to convert
these indices into $\teff$ and $A_V$. We also show that this technique
is not significantly affected by optical excesses from mass accretion,
typical of young stars and BDs. In Section \ref{section:HRD} we derive
the new HRD for the ONC, which now includes $\sim 1800$ sources, reaching
masses as low as $0.02~M_\odot$. We study the completeness as a function of
$\teff$ and bolometric luminosity ($L_{\rm bol}$)
accounting for photometric detection as well as
differential reddening; we detect a population of candidate background
contaminants, which appears well separated from the ONC members in the
HRD, at lower luminosity. After excluding the contaminants, in Section
\ref{section:IMF} we derive and discuss the IMF of the ONC.

\section{The data}
\label{section:the_data}
\subsection{Observations}
\label{subsection:observations}

Observations were carried out with the Wide Field Imager (WFI), a
focal reducer-type camera mounted at the Cassegrain focus of the 2.2-m
MPG/ESO telescope at La Silla. The field of view (FOV) of the camera is
$\sim32\arcmin \times33\arcmin$, allowing us to cover the entire ONC with
one pointing.

\begin{figure}
\epsscale{1.1}
\plotone{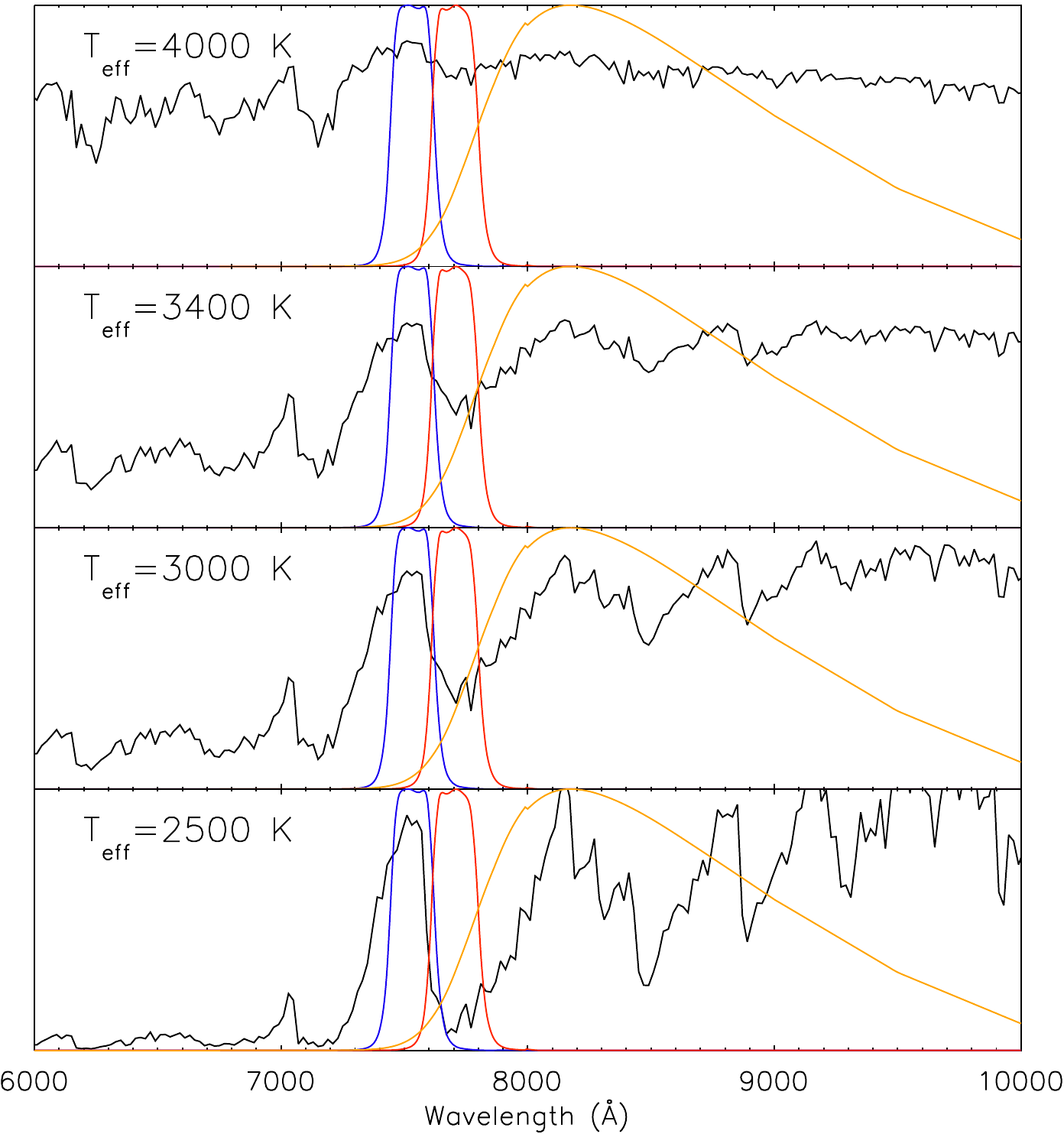}
\caption{The band profiles of the 3 WFI filters used in this work (from left to right: 753, 770 and $I-band$) overlaid to spectra of very low-mass stars (4000~K to 3000~K) and BDs (3000~K to 2500~K) from \citet{allartio}.
\label{fig:spectra}}
\end{figure}

This optical imager offers a particularly large set of broad-, medium-,
and narrow-band filters along the entire wavelength range of its CCD
sensitivity ($\sim$3000 to 10000$\AA$). We selected two WFI medium
band filters which provide the best sampling of a deep absorption band
at $\lambda\sim7700\AA$. These are the MB$\sharp$753/18-ESO848 and the
MB$\sharp$770/19-ESO849 filters, which hereafter will be simply referred
to as \emph{753} and \emph{770}. Figure \ref{fig:spectra} shows the
transmission profiles of these filters overlayed to synthetic spectra from
\citet{allartio} of decreasing temperature. The $753$ band is centered
on the photospheric continuum, while the 770 is completely covering the
TiO absorption band whose strength increases with decreasing $T_{\rm
eff}$. We also used a broad $I$-band filter, the BB$\sharp$I/203-ESO879,
to sample the stellar flux at redder wavelengths, in order to evaluate
the photospheric reddening from dust extinction. This I-band filter is
also used in our previous photometric studies of the ONC (Paper~I and II).
Table \ref{tab:bands} summarizes the characteristics of the 3 filters.

\begin{deluxetable}{rrrrr}
\tablecaption{Characteristics of the filters used for our WFI imaging \label{tab:bands}}
\tablehead{\colhead{band} & \colhead{$\lambda$ central} & \colhead{FWHM} & \colhead{E.W.\tablenotemark{a}} & \colhead{Vega zero-point} \\
\colhead{} & \colhead{($\AA$)} & \colhead{($\AA$)} & \colhead{($\AA$)} & \colhead{(erg s$^{-1}$ cm$^{-2}$ $\AA^{-1}$)} }
\startdata
753 & 7536.2 & 184.1 & 191.4 & 1.368$\cdot 10^{-9}$ \\
770 & 7707.9 & 195.2 & 201.3 & 1.275$\cdot 10^{-9}$ \\
I & 8620.5 & 1354.9 & 1431.2 & 9.500$\cdot 10^{-10}$
\enddata
\tablenotetext{a}{Equivalent Width}
\end{deluxetable}

The WFI
imaging was carried out during 5 nights of December 2010. On each night,
observations were performed in all the three bands together
so as to minimize the impact of any longer timescale
stellar variability. A standard dithering pattern
has been applied, in order to cover the gaps between the 8 CCD chips of
WFI and allow for cosmic ray and bad pixel removal.
We did not carry out short exposures to avoid saturation
for the bright sources, since the bright end of the ONC has been studied
well in detail by us in our previous works.
Dark, bias
and flat field exposures were obtained before and after each night. We did
not observe standard fields for photometric calibration, given the particular
filters used in this program and the practicality of performing relative
photometric calibration exploiting our large pre-existing data set on Orion.

\begin{deluxetable}{rrrr}
\tablecaption{WFI observations \label{tab:observations}}
\tablehead{\colhead{band} & \colhead{exposure time} & \colhead{n. of exposures} & \colhead{total exposure time}  \\
\colhead{} & \colhead{($s$)} & \colhead{} & \colhead{(s)}  }
\startdata
753 & 360 & 23 & 9,360  \\
770 & 500 & 28 & 14,000 \\
$I$   & 280 & 12 & 3,360
\enddata
\end{deluxetable}


\subsection{Data reduction}
\label{subsection:data_reduction}

Images were processed using the ESO/MVM (Multi-Vision Model) {\em vers.}
1.3.5 package \citep{vandame04}. This software automatically performs all
common image reduction steps (e.g., correction for bias and flat fields),
and the absolute astrometric calibration of the individual exposures to
be merged. This is accomplished by cross-correlating the images with a
reference astrometric catalog; we utilized to this purpose our previous WFI
photometric catalog presented in Paper~I. The final product of our image
reduction is 3 images, one for each filter, obtained by merging all the
individual exposures. Table \ref{tab:observations} summarizes the total
exposures times; these about 1 hour for the broad band, and about 2.5 and 4 hours for the two medium bands,

Photometry, both aperture and point-spread-function (PSF), was performed
using the Daophot II package
\citep{stetson87}. We computed the PSF individually for every image, starting
from a coadded sample of few hundreds bright and unsaturated sources,
and then refining these with an iterative sigma-clipping
method to reject PSF stars with poor $\chi^2$. Photometry was computed on
all the sources brighter than 3$\sigma$ above the local sky background
noise. In order
to eliminate the contamination from spurious detections---a potentially
significant fraction of the total detections, given the strong variability
of the nebular background, and the low threshold for source detection---we
rejected sources not present in any of our existing deep photometric
catalogs of the ONC, both ground based (WFI $UBVI$, Paper~I; CTIO/ISPI $JHK$, see
\citealt{robberto2010}) or from the HST photometry (ACS $BVIZ$, WFPC2
$UBI$, see \citealt{robberto2005treasury}). In particular, the significant
depth of the infrared and HST data, over a FOV equal or larger than that
of the observations described here, guarantees that
we are not excluding any real sources in this study.

The photometry has been calibrated to the VegaMag system using our
previous WFI photometric catalog. For the $I$-band, we simply determined
the photometric offset with respect to our previous photometry in the
same filter.
For the two medium bands, we used synthetic photometry. We considered the atmosphere models of \citet{allartio} as
empirically calibrated by us in Paper~II, and computed the two medium-band colors $(MB-I)$ as a function of $\teff$.
Then we considered the actual sources detected in all 3 bands and for which we have $\teff$ and $A_V$ from Paper II, and
plotted the measured, extinction-corrected ($MB_{\rm instrumental}-I_{calibrated}$) colors against $\teff$.
The systematic difference between these colors (considered individually) and the synthetic ones is the offset needed to calibrate the medium bands.
Specifically, we limited to the stars with $T_{\rm eff}>4000$~K, range where the possible inaccuracies of the atmosphere models are very modest.


\subsection{Completeness}
\label{subsection:photometric_completeness}
\begin{figure}
\epsscale{1.1}
\plotone{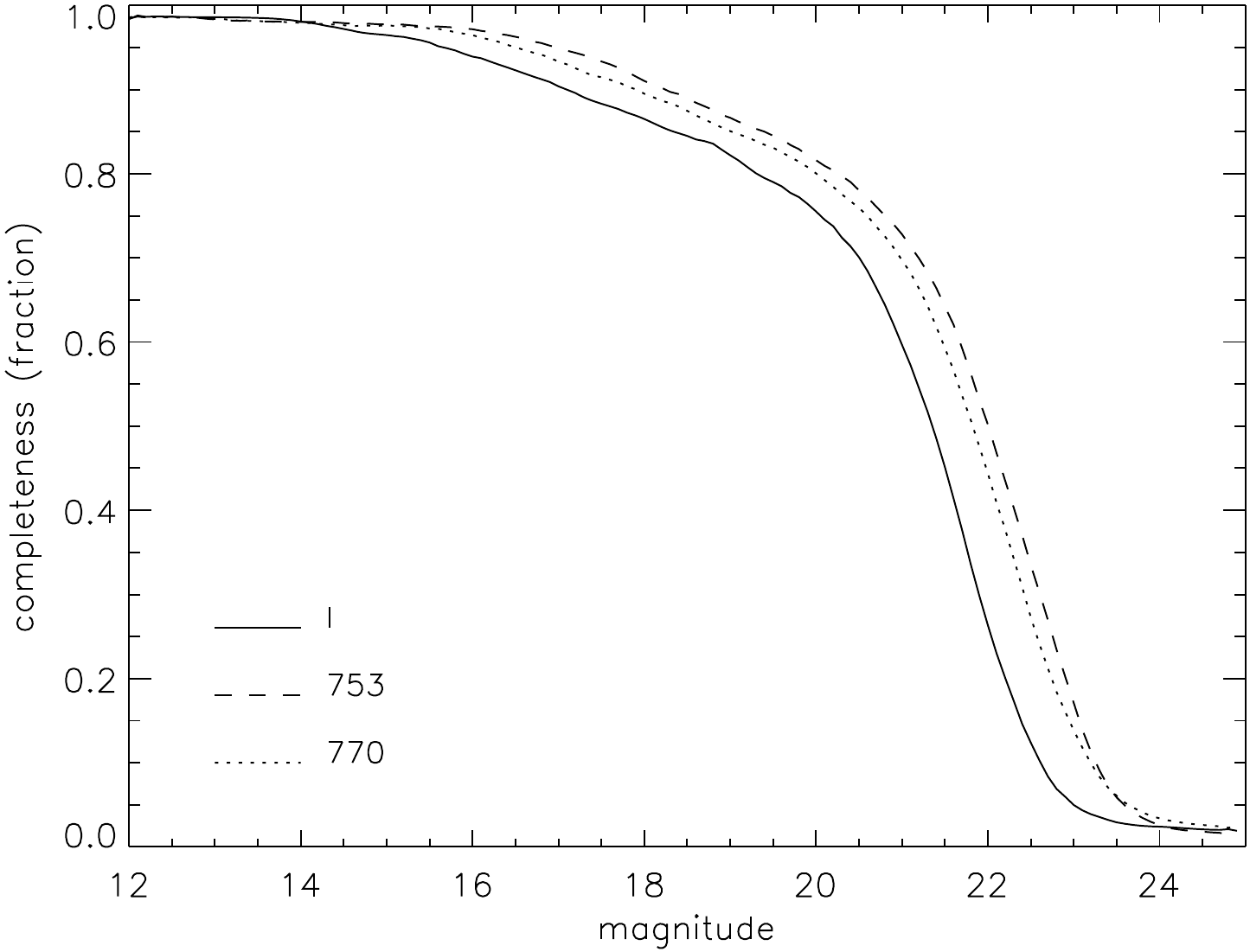}
\caption{Photometric completeness functions for our three filters. \label{fig:plot_completeness_mag}}
\end{figure}

We derive the completeness functions for our WFI photometry with an artificial star test, performed separately for each WFI filter. Artificial stars of different magnitude, are added on the reduced WFI images; then, with the same technique used to detect and extract photometry on the real stars, these artificial stars are recovered or missed depending on their flux, local sky noise, crowding, etc. The fraction of recovered stars against the total, as a function of magnitude, provides the completeness.

In our ONC imaging, the completeness is spatially variable, being strongly affected by the nebular background, which is brighter at the center where the stellar density is higher. This correlation, if not accounted for, may bias the derived completeness. Therefore, we first derive the spatial, projected, stellar density distribution of the ONC. This is obtained by merging all catalogs at our disposal, in particular the deep ISPI $JHK$ photometry and the HST/ACS data. Then, for every magnitude bin, the artificial stars are generated with random positions drawn from the 2 dimensional density distribution.

The derived completeness functions are shown in Figure \ref{fig:plot_completeness_mag}. In all cases, these curves show a characteristic trend: besides the typical steep cutoff at faint luminosities which traces the detection limit, the completeness shows a slow decrease below 1 at bright luminosities. This is due to the luminous nebular emission and significant crowding of the central part of the ONC, where the detection limit is significantly poorer. In the outer regions of the ONC, we are able to detect sources all the way down to $m\sim 22.5$~mag in $I$ band (see Figure \ref{fig:CMD}), and to $m \sim 23.5$~mag in $753$ and $770$.

\subsection{Photometry results}

\begin{figure}
\epsscale{1.1}
\plotone{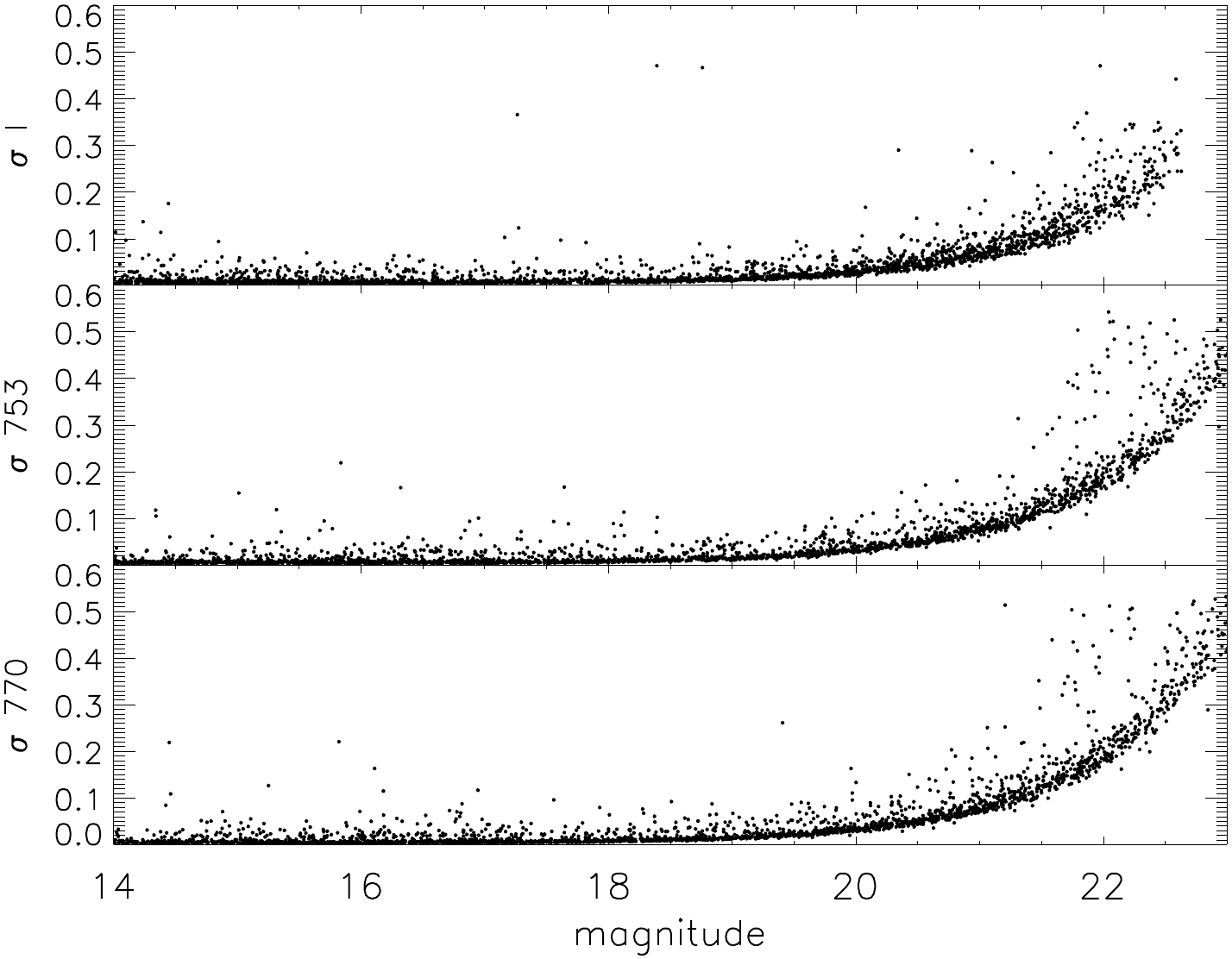}
\caption{Photometric errors as a function of magnitude, for the 3 photometric bands \label{fig:photometric_errors}}
\end{figure}
\begin{figure}
\epsscale{1.1}
\plotone{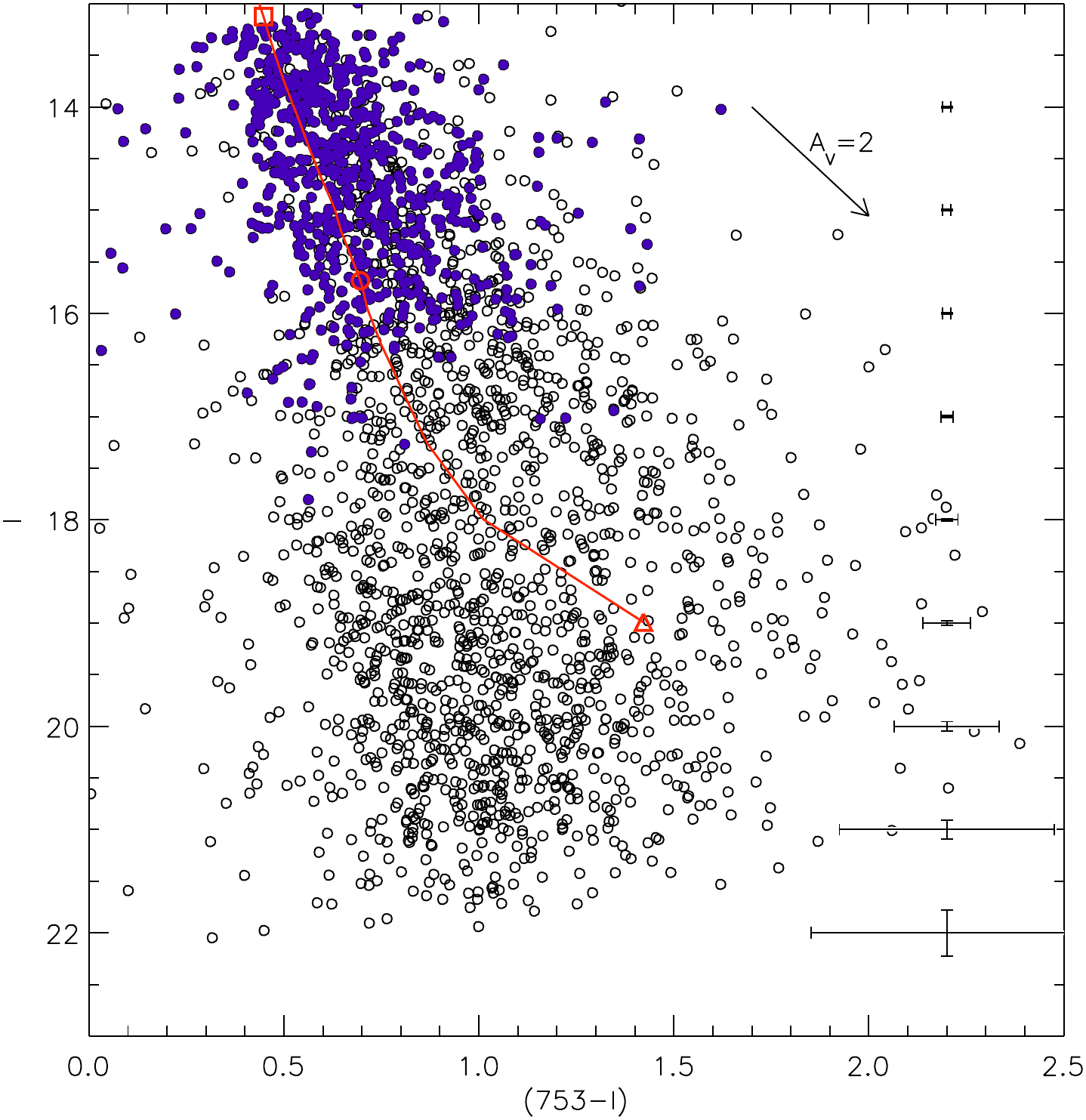}
\caption{Open circles: the derived $I$ vs $753-I$ CMD including only the sources measured in all the 3 bands (i.e. also in $770$). The filled blue circles indicate the sources included in our previous study of the stellar population of the ONC, based on WFI photometry and spectroscopy down to the H-burning limit (Paper~II). The average photometric errors as a function of $I$-band are indicated to the right. The red line represents a 1~Myr \citet{bcah98} isochrone; the square, circle and triangle symbols along the line mark masses of 0.4~M$_\odot$, 0.08~M$_\odot$ and 0.02~M$_\odot$. The arrow denotes the reddening vector for $A_V=2$ assuming the \citet{cardelli} reddening law for $R_V=3.1$. \label{fig:CMD}}
\end{figure}

\begin{figure*}
\epsscale{0.8}
\begin{center}
\plotone{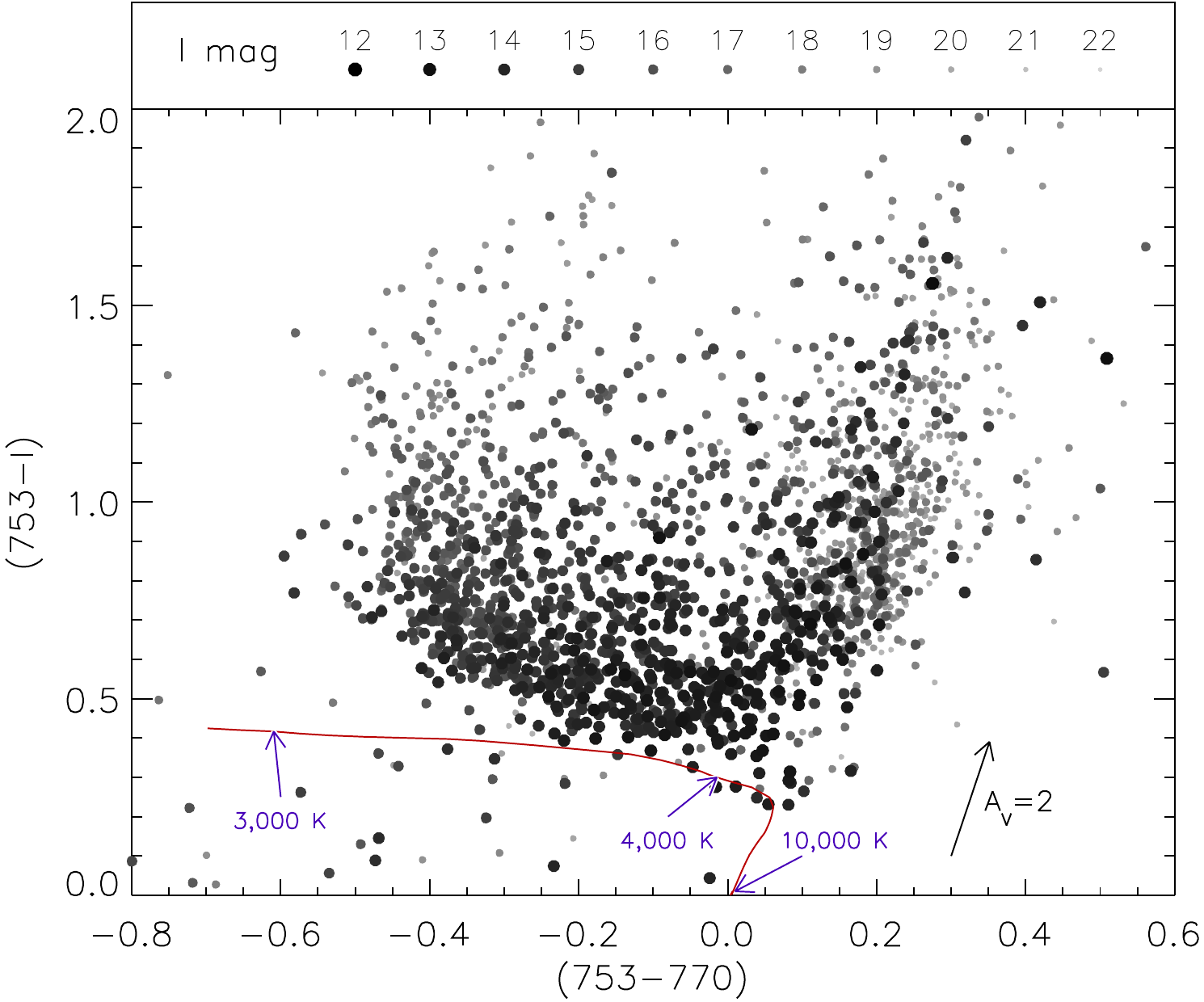}
\end{center}
\caption{$(753-I)$ vs. $(753-770)$ color-color diagram from our
photometry. The color and size of the dots indicates the I-band luminosity
of each source, as indicated. The black arrow indicates the direction of
the reddening vector for 2~mag of extinction in $V-$band, assuming
the reddening law of \citet{cardelli} for $R_V=3.1$.
For reference, the thick red line
is the empirical isochrone derived in Paper~II converted in this
color-color plane using the atmosphere models of
\citet{allartio}. For guidance,
the origin (0,0) corresponds to 10,000~K, and the turn-over in $(753-770)$
corresponds to $\teff\sim4000$~K, i.e. between K and M spectral classes.
\label{fig:color_color_diagram}}
\end{figure*}

The final catalog includes 2474 sources with photometry in all 3
bands. The photometric errors as a function of magnitude are shown in
Figure \ref{fig:photometric_errors}. Figure \ref{fig:CMD} shows our new
CMD, highlighting the significant improvement in depth with respect to our
previous work (Paper~II). In particular,
our photometry covers a very large range of stellar luminosities,
spanning $\sim 9$~mag in $I$-band, and the new observations extend
$\sim$5~mag deeper than in the previous work, reaching $I\simeq 22$,
which corresponds to young sources with masses $<0.02$~M$_\odot$ at 1~Myr
(see Figure \ref{fig:CMD}).
In Figure \ref{fig:CMD} we also observe among the faintest sources
a concentration of objects with $I\gtrsim18$ which, as we show below,
comprise a distant, background stellar population that is
detectable despite the high extinction through the molecular cloud
because of the depth of our observations.

In Figure \ref{fig:color_color_diagram} we present the
$(753-I)$ vs $(753-770)$ color-color diagram,
which forms the basis of our analysis
in all that follows.  The most salient feature of this diagram is the
inflection point near (0.05,0.25), from which the stars extend in two
different directions. By comparison with the orientations of the synthetic
isochrone and the extinction vector, we see that this inflection in the
diagram essentially separates the stars into a group
(extending to the left) that stretches mostly parallel to the isochrone
and perpendicular to the reddening vector, and a second group
(extending to the upper right) that stretches parallel to both
the isochrone and the reddening vector.
The inflection in the diagram occurs at $\teff \approx 4000$~K.

For stars with $\teff\lesssim 4000$~K (i.e. spectral types $>$M0),
the $(753-770)$ color is both highly sensitive to $\teff$,
becoming rapidly bluer (more negative) as $\teff$ decreases
(see also Figure \ref{fig:spectra}),
and largely insensitive to $A_V$ because of
the small wavelength separation between the 753 and 770 bands.
In our analysis below we exploit this property of the
diagram to break the degeneracy between $\teff$ and $A_V$ for these
very cool stars, which has previously been
one of the most difficult challenges in the derivation of
stellar parameters from optical broadband photometry for M-stars and BDs
(\citealt{hillenbrand97}; Paper~II).
We note here that the synthetic isochrone clearly does not
correctly reproduce the slight upward curvature of the stars as one
moves to the coolest $\teff$, but because these cool stars nonetheless
extend perpendicular to the reddening vector, we demonstrate below that
the systematic error in the $(753-I)$ colors predicted by the
synthetic isochrones can be readily
calibrated out for $\teff\lesssim 4000$~K.
We note also that the effects of contamination by faint background
sources is mitigated in this region of the diagram, particularly for
the coolest objects of interest
($\teff\lesssim2900$~K, i.e., in the BDs regime),
since background BDs are too intrinsically faint to penetrate the
very high $A_V$ of the cloud behind the ONC.
As we show in Section \ref{section:HRD}, the relatively few remaining
background contaminants in this region of the diagram (mostly red giants)
can be filtered out on the basis of their luminosity.

These desirable features of the
$(753-I)$ vs $(753-770)$ color-color diagram
break down for warmer stars with $T_{\rm eff}>5000$~K.
For these warmer stars the colors are largely degenerate in $\teff$ and
moreover the isochrone and the reddening vector are parallel.
For these stars we must therefore rely
on previous observations by us and from the
literature to determine $\teff$ and $A_V$.
In addition, there is significant background contamination in this
region of the diagram;
as we show below, while the brightest of these stars are bona fide ONC
members, the faintest are mostly background stars.
Finally, we note that the apparent lack of low-$A_V$ stars for
$\teff\gtrsim 4000$~K is the result of these stars being
saturated in our deep images.
Such stars are indeed present in the ONC, and we include them in our
analysis from our own previous work (Paper~I and II).




\section{Analysis}
\label{section:analysis}

\begin{figure}
\epsscale{1.1}
\plotone{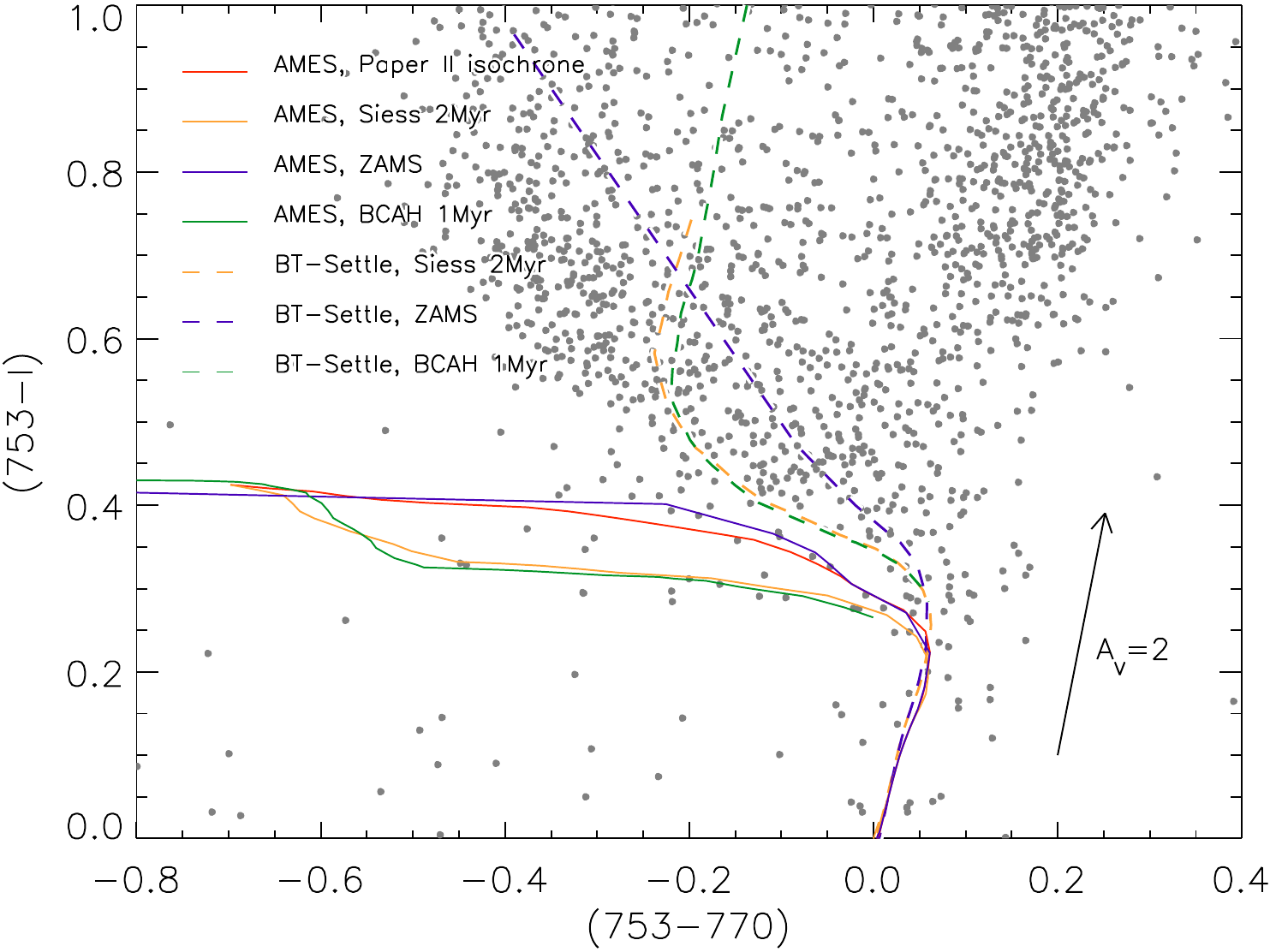}
\caption{The color-color diagram of Figure \ref{fig:color_color_diagram}, with different synthetic isochrones overplotted. In particular: a \citet{siess2000} 2~Myr isochrone; a \citet{bcah98} 1~Myr isochrone; a \citet{siess2000} zero-age main sequence. Thick lines are the result from synthetic photometry on the AMES-MT grid of spectra from \citet{allartio}, while dashed line are computed using the BT-Settle grid from \citet{allard2010}. The thick red line is computed using the empirical calibration of the AMES-MT models defined in our Paper II to match the broad band $BVI$ colors of the ONC. The empirical line is actually closer to the predicted ZAMS relation than the predicted 1-2~Myr isochrones, illustrating the current lack of agreement between evolutionary and atmosphere models and young star photometry (see text).  \label{fig:color_color_models}}
\end{figure}
\begin{figure}
\epsscale{1.1}
\plotone{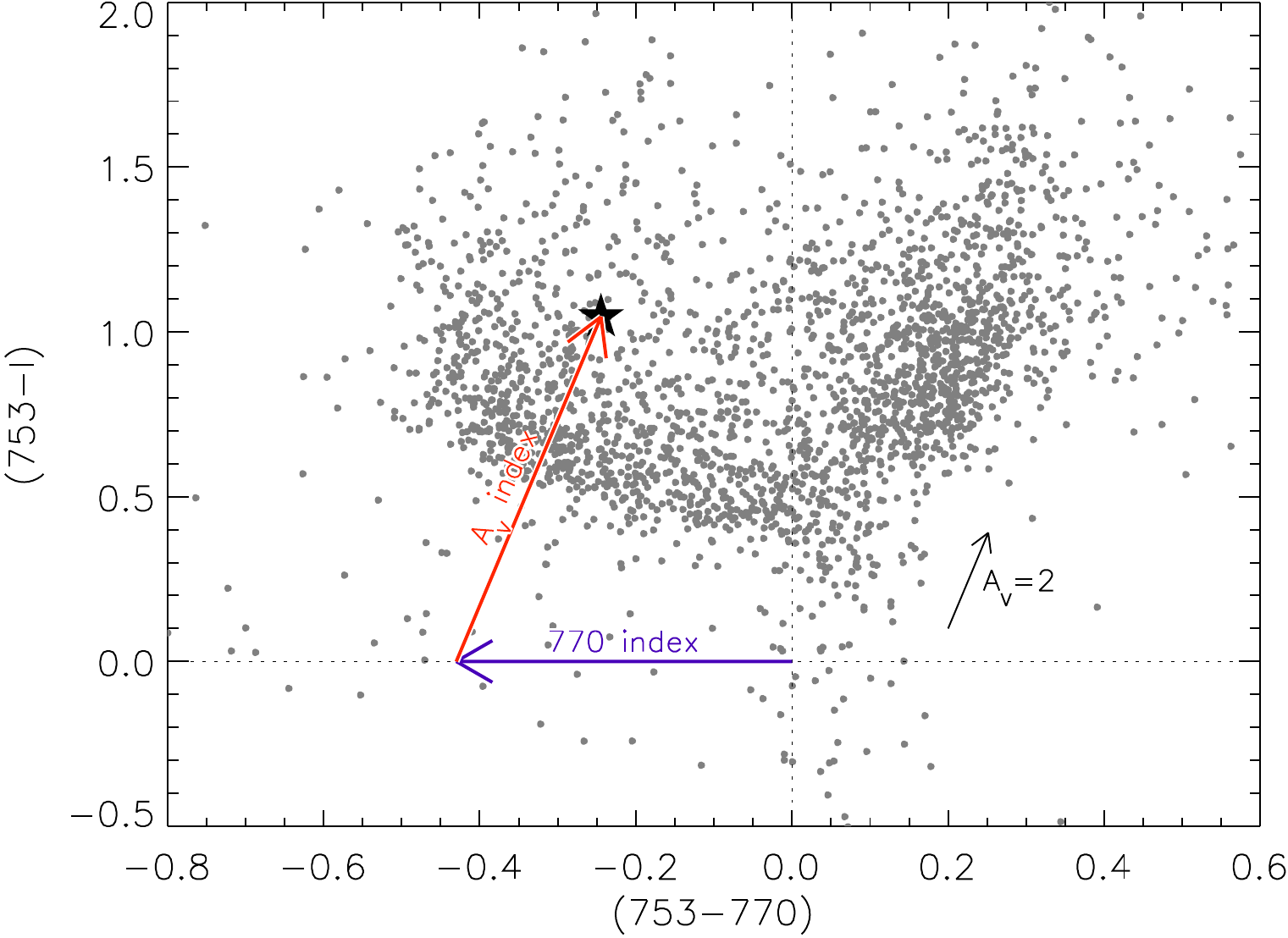}
\caption{Conceptual scheme of the new coordinate system we introduce to measure $\teff$ and $A_V$ of the stars from our color-color diagram: the $\teff$ dependent [$770~index$], and the reddening dependent $A_V$ index. See text. \label{fig:color_color_indexes}}
\end{figure}

In order to derive the stellar parameters of very low-mass stars
(VLMSs) and BDs in the ONC, we focus our analysis on the left-hand part of
the color-color diagram of Figure \ref{fig:color_color_diagram}, where the
$(753-770)$ color anti-correlates with $\teff$.

Our basic methodology is to
project each source from its observed position in the color-color
diagram back along the reddening vector to the theoretical isochrone.
The choice of theoretical isochrone is largely arbitrary because
none of the available isochrones exactly reproduce the observed
distribution of the stars in the diagram, and consequently we must
empirically recalibrate the isochrone based on the observed distribution.
Our assumption is that, whichever fiducial isochrone is adopted, the
non-reddened stars should lie on the isochrone, and the reddened stars
displaced away from it along the reddening vector.
For example, the isochrone shown in Figure \ref{fig:color_color_diagram}
computed using the synthetic spectra of \citet{allartio}
predicts a much bluer $(753-I)$ index than our data
for low $(753-770)$. If the model were correct, this would imply
an unrealistic dependence of the average $A_V$ on $\teff$,
in the sense that the minimum measured extinction (i.e. the distance,
along the reddening vector, between the isochrone and the stars) increases
monotonically with decreasing $\teff$.

Different synthetic isochrones can lead to more realistic
predictions of colors.  Figure \ref{fig:color_color_models} shows our
color-color diagram with a number of synthetic isochrones, computed
using both the AMES-MT spectra of \citet{allartio} and the most recent
grid---the BT-Settle models---from \citet{allard2010}. We also show an
isochrone (red line) derived using the \citet{allartio} atmospheres
as recalibrated by us in Paper~II. That calibration, appropriate for the ONC, involved an empirical
constraint on the stellar $\log g$ to produce a synthetic isochrone that
matched the observed broadband colors of the ONC in $BVI$.
It is clear that all synthetic isochrones, including that which we used
in Paper~II, are incompatible with the data in this new medium-band
color-color plane.
AMES-MT consistently predicts a systematically lower $(753-I)$ at low
$\teff$, and BT-Settle predicts either a turn-over in $(753-770)$ for
low $\log g$ (young isochrones) or $(753-I)$ colors that are too red for
high $\log g$ (the ZAMS model).
Evidently, current synthetic models, even those that have been empirically
calibrated by us to match the broadband colors of young low-mass stars,
are unable to correctly reproduce the molecular feature at $\lambda\sim7700$\AA\
and/or the photospheric continuum level at $\lambda\sim7500$\AA\ which
form the basis of our new medium-band color-color diagram technique.  As a
consequence, we must empirically define an isochrone representing both
the photospheric colors in our bands and the $\teff$ scale along it.

To proceed, we introduce a new set of coordinates in the
color-color plane of Figure \ref{fig:color_color_diagram}.
We define the $[A_V~index]$ as the distance that a given star is
displaced, along the reddening vector, from the $x-$axis of the
color-color diagram, measured in units of $A_V$:
\begin{eqnarray}
\label{equation:Avindex}
[A_V~index]= & \bigg(m_{753}-I\bigg)\bigg/\bigg(\frac{A_{753}}{A_V}-\frac{A_I}{A_V}\bigg)
\end{eqnarray}
\noindent where $A_{753}/A_V=0.678$, $A_{770}/A_V=0.652$ and
$A_{I}/A_V=0.533$ (values computed from the \citet{cardelli} reddening law
for $R_V=3.1$). We choose the reddening parameter $R_V=3.1$ instead of a
higher value (e.g, 5.5, \citealt{costero70}) because, as we found in Paper~II,
this value better explains the observed broadband colors of the ONC members.
We similarly define the $[770~index]$ as the  projection of
a given star, along
the direction of the reddening vector, onto the x-axis [i.e. $(753-I)=0$]:
\begin{eqnarray}
\label{equation:770index}
[770~index]= & m_{753}\cdot\bigg[1-\big(\frac{A_{753}}{A_V}-\frac{A_{770}}{A_V}\big)/\big(\frac{A_{753}}{A_V}-\frac{A_{I}}{A_V}\big)\bigg] + \nonumber  \\
& + I\cdot\bigg[\big(\frac{A_{753}}{A_V}-\frac{A_{770}}{A_V}\big)/\big(\frac{A_{753}}{A_V}-\frac{A_{I}}{A_V}\big)\bigg] + \nonumber \\
& - m_{770}
\end{eqnarray}
By definition, the $[770~index]$ is reddening independent,
and so it depends only on $\teff$.
Figure \ref{fig:color_color_indexes} shows a schematic
representation of the two indices in the color-color diagram.

\subsection{Derivation of $\teff$}
\label{subsection:teff_calibration}
Because the currently available synthetic spectra are unable to accurately
predict the stellar colors in our medium-band color-color diagram,
we must derive an empirical transformation between the
$[770~index]$ and $\teff$. To this end we utilize the sample of ONC
stars with directly determined (i.e. spectroscopic) spectral types.
With the appropriate transformation between the $[770~index]$ and $\teff$,
we will be able to reliably determine $\teff$
of all the other members from their measured [$770~index$].
We have assembled all available optical spectral types from our Paper~II
(for stellar-mass objects)
and NIR spectral types (for substellar-mass objects) from
\citet{riddick2007} and \citet{slesnick04}.
The spectral types were converted to $\teff$ using the
temperature scale of \citet{luhman2000}, as we did in Paper~II.

\begin{figure}
\epsscale{1.1}
\plotone{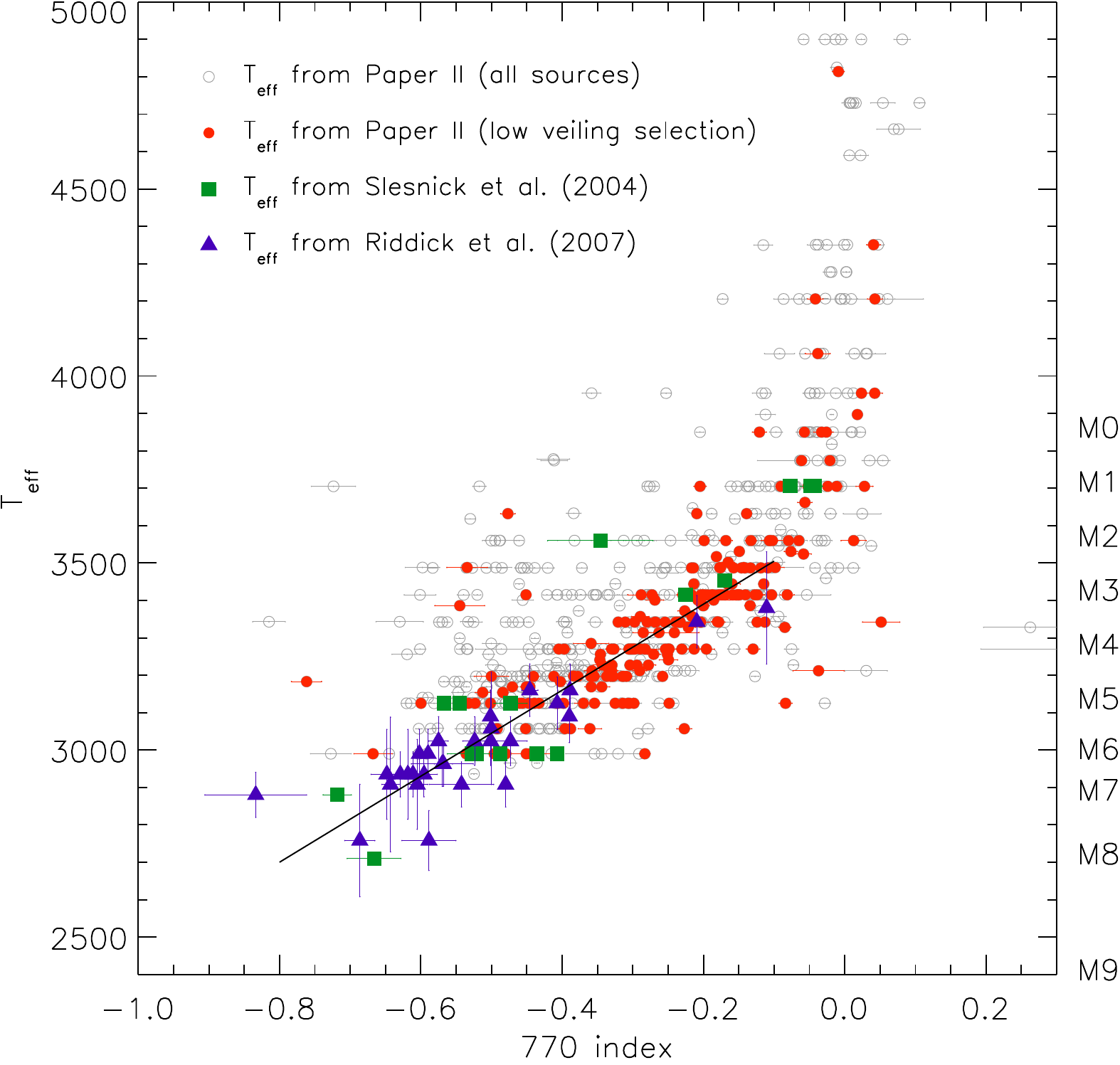}
\caption{Derived relation between the measured, reddening independent,
photometric $[770~index]$, and spectroscopically determined stellar
temperature. The values of $\teff$ for the stellar mass regime ($\teff\gtrsim
3000$~K) are from the spectral types collected in our Paper~II. For these
sources, we identify a sample of candidate weak accretors, showing low
H$\alpha$ emission (E.W.$<10\AA$) and no significant veiling ($\log L_{\rm
accr}/L_{\rm tot}<-1.5$), in order to more cleanly determine the true
relationship between the photometric $[770~index]$ and $\teff$.
Spectroscopic $\teff$ for substellar masses are
from \citet{riddick2007} and \citet{slesnick04}. See text for details.
\label{fig:index_teff}}
\end{figure}

Figure \ref{fig:index_teff} shows the resulting relationship between
our measured [$770~index$] and the spectroscopically determined
$\teff$. As expected, for late spectral subtypes
the $[770~index]$ becomes bluer (i.e. becomes more negative)
with $\teff$. This approximately linear dependence holds from
$\teff\sim3500$~K ($\sim$M2.5) down to the coldest available spectral types
for BDs, and down to our photometric lower limit ($[770~index]\simeq0.8$).

Perhaps not surprisingly, we found that the scatter in the
[$770~index$]--$\teff$ relation could be significantly reduced if we
consider only sources that do not show strong spectroscopic
evidence of accretion. We selected the ``non-accretors" on the basis
of low H$\alpha$ emission (E.W.$<10\AA$) from Paper~I and low optical veiling
(low $B-$band excess emission) from Paper~II.  Figure \ref{fig:index_teff}
shows that these non-accretors form a relatively tight sequence,
while the rest of the stars (the accretors) are much more scattered in
the diagram.  The accretors are also systematically shifted to higher
$\teff$ and/or lower [$770~index$], indicating that accretion alters
the position of the sources in the diagram, and indeed below we exploit
this feature to quantify the degree to which accretion may affect our
determination of $\teff$.

Using only the sample of non-accretors,
we derive an empirical linear relation between $\teff$ and the $[770~index]$:
\begin{equation}
\label{equation:teff_vs_index}
\teff = 3620~{\rm K}+[770~index] \cdot1150~{\rm K}
\end{equation}
\noindent which is valid for $[770~index]<-0.1$. The standard deviation of
the data relative to this relation is $\sim 100$~K, roughly corresponding
to one-half of a spectral subtype. This is comparable to the overall
uncertainty in our spectroscopically determined $\teff$ from Paper~II,
suggesting that this scatter
is principally due to
uncertainty in the (spectroscopic) spectral types, and therefore any
additional uncertainty introduced through the $[770~index]$--$\teff$
transformation is negligible.

\subsection{Influence of accretion excess on the $\teff$ estimation.}
\label{subsection:accretion}


The analysis methodology utilizing our medium-band color-color diagram
is predicated on the assumption that the stellar colors in our
photometric bands depend only on $\teff$ and $A_V$,
permitting us to establish an empirical relationship between the
our [$770~index$] and $\teff$ (Equation \ref{equation:teff_vs_index}).
However, the colors of young stars can be affected by
accretion, whereby infalling matter from
the circumstellar disk onto the stellar surface leads to a flux excess,
strongest in the UV and in specific emission lines, but also to a
smaller extent in the continuum at optical wavelengths.
The latter is typically characterized
by a heated photosphere covering $\sim$1\% of the stellar surface and with
$\teff\sim6000-8000$~K, therefore peaking in
the B-band \citep{calvetgullbring98}. Since our photometry uses filters at the red end of the
optical spectrum, and there are no emission lines within the wavelength
range covered by our filters, we expect that accretion should induce
only a modest alteration of our measured colors.

\begin{figure}
\epsscale{1.1}
\plotone{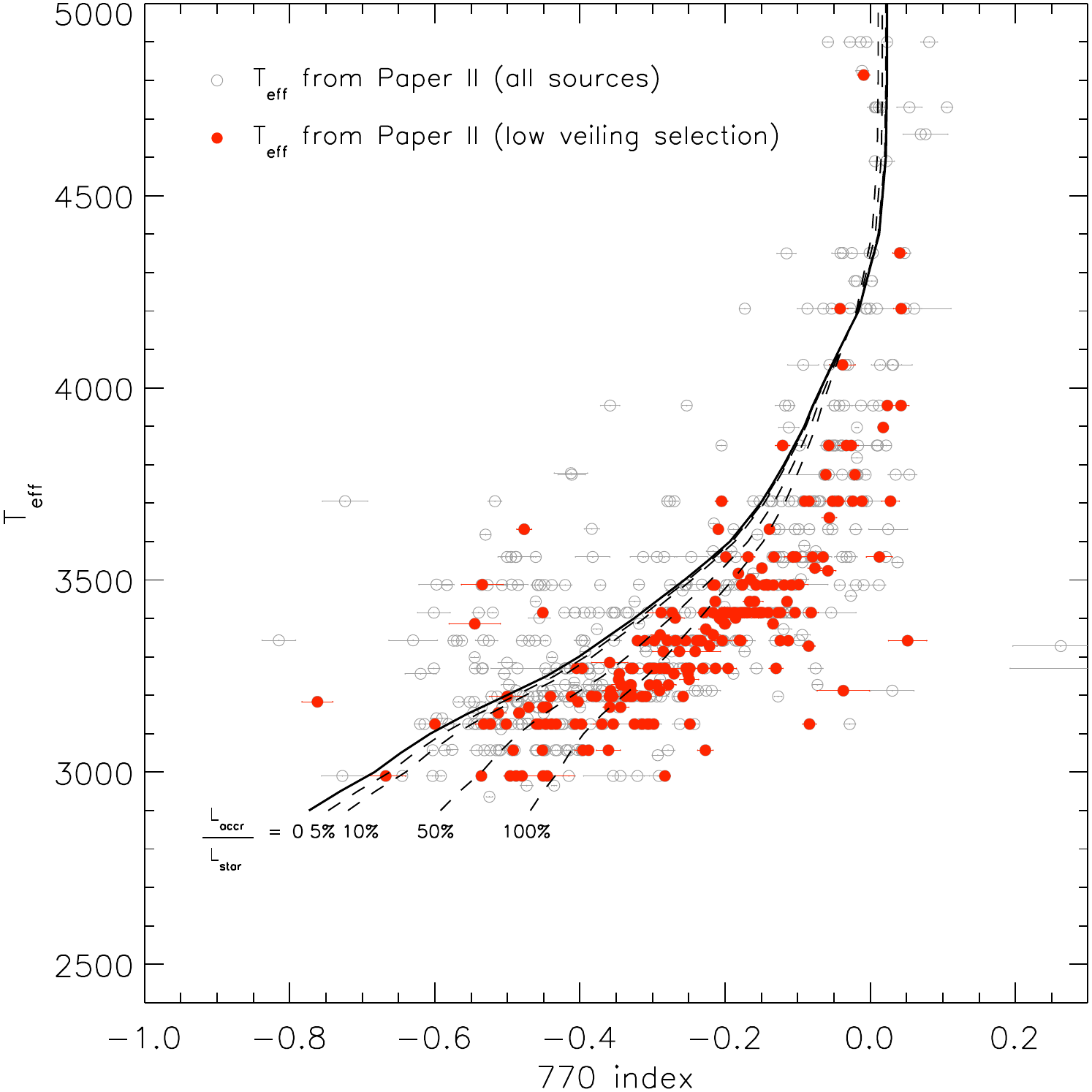}
\caption{Same as Figure \ref{fig:index_teff}, showing the displacement in the reddening-independent $[770~index]$ obtained by adding an amount of accretion luminosity to the star. The thick line is the isochrone from Paper~II, the same model shown also in Figure \ref{fig:color_color_diagram}. The dashed line are the same model, after the addition of a given amount of accretion luminosity to the each synthetic spectrum, for different values of $\log L_{\rm accr}/L_{\rm tot}.$ Note that the fact that the red dots are low accretion sources, whereas the model requires a large value of $L_{\rm accr}$ to predict their colors, implies that the base isochrone needs adjustments (see text). \label{fig:index_teff_accretion}}
\end{figure}
\begin{figure}
\epsscale{1.1}
\plotone{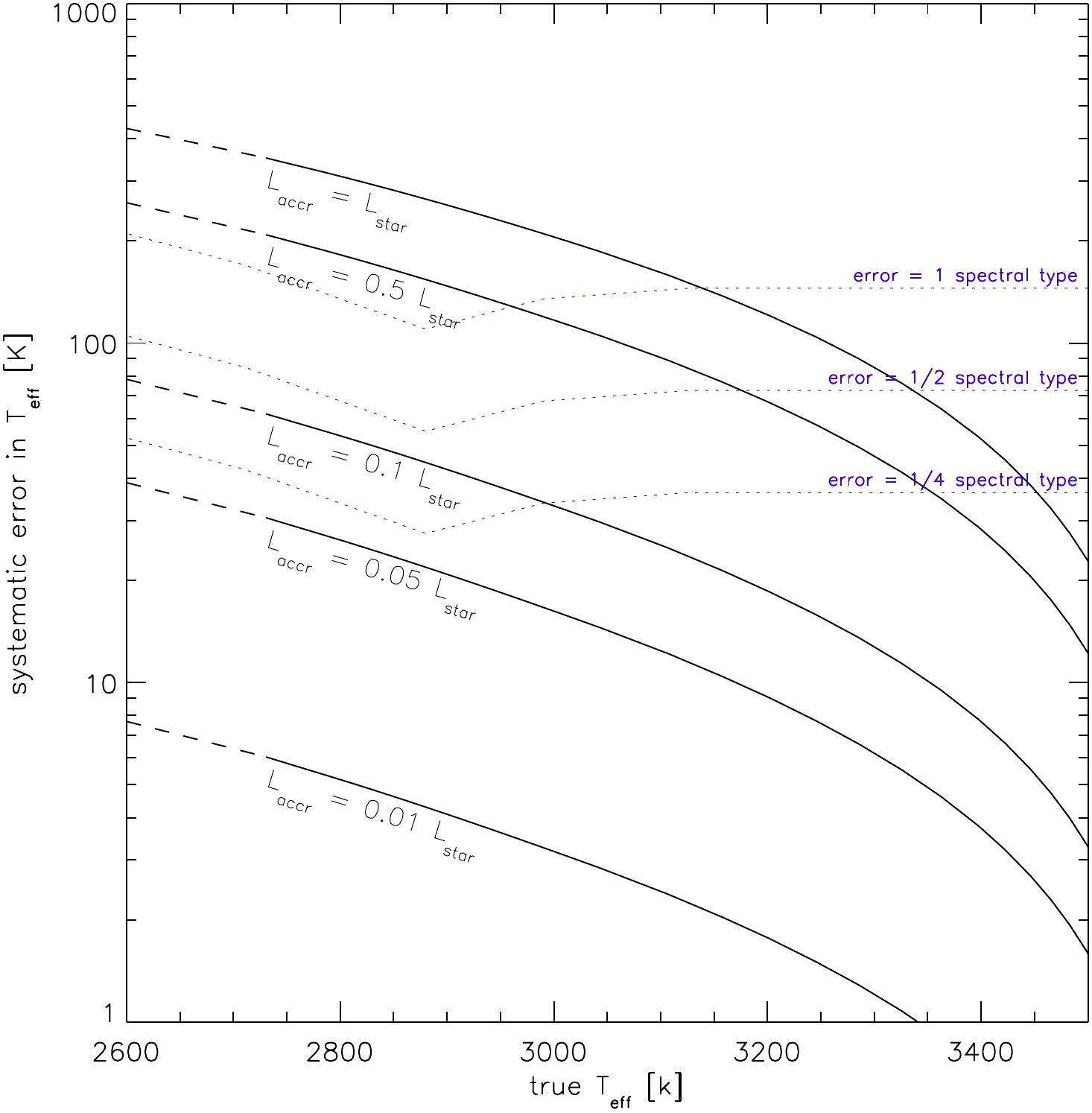}
\caption{Computation of the offset on the derived $\teff$, as a function of temperature, produced by neglecting accretion veiling. The solid lines represent the result for different accretion luminosities. The dotted lines represent the uncertainty in $\teff$ associated to un uncertainty in the spectral type of 1 subtype, half subtype and 1/4 of a subtype. \label{fig:accretion_bias}}
\end{figure}


We investigate this hypothesis in a quantitative way, calculating the
shift in [$770~index$], at constant $\teff$, obtained by contaminating
the photospheric fluxes with varying degrees of
accretion luminosity.  As in Paper~II,
where we performed a similar modeling to derive the change in the $BVI$
colors due to mass accretion, we first model a typical accretion spectrum,
then we add it with increasing relative flux fraction on synthetic stellar
spectra of different $\teff$. Finally, by means of synthetic photometry
as above, we compute the resulting changes in the $[770~index]$
and in turn the error induced in the inferred $\teff$.  Specifically,
we simulated a typical accretion spectrum using the \textsc{Cloudy}
photoionization code. This is similar to that we used in Paper~II,
but has been further refined to provide a more realistic SED in the
ultraviolet range (Manara et al. 2011, \emph{in prep.}).  We use the
synthetic spectra that we used in Paper~II--namely the empirically
calibrated AMES-MT model from \citet{allartio}---and for every $\teff$
we add a component of our accretion spectrum to the photospheric spectra,
varying the ratio $L_{\rm accr}/L_{\rm star}$.

The result is shown in Figure \ref{fig:index_teff_accretion},
from which two main results are evident:
First, contaminating the stellar flux with accretion excess shifts the
[$770~index$] to higher values, and the effect is more prominent for cooler
$\teff$. This is intuitive considering that this index traces
the ratio between the flux outside ($\lambda\sim7530$\AA) and inside
($\lambda\sim7700$\AA) the TiO molecular feature. The accretion-induced
continuum excess partially ``fills in'' the TiO absorption feature,
leading to higher (less negative) values of [$770~index$].
Second, for typical values of accretion luminosities in a PMS cluster
($L_{\rm accr} \lesssim L_{\rm star}$) the shift in the [$770~index$]
is negligible.

From Figure \ref{fig:index_teff_accretion} it also appears that the synthetic
isochrone does not quite follow our data. This is not surprising, given
that, as discussed in Section \ref{section:the_data}, this model provides
a somewhat inaccurate estimate of the stellar fluxes in our medium
bands (see Figure \ref{fig:color_color_models}). Therefore, in order
to obtain a correct, quantitative assessment of the accretion-induced
shift in the [$770~index$], we must recalibrate the results shown in
Figure \ref{fig:index_teff_accretion}.  We proceed as follows. The
systematic error in the depth of the TiO feature as a function of $\teff$
can be treated as a simple systematic error in the $\teff$ scale.
For example, for the lowest $\teff$ modeled in Figure
\ref{fig:index_teff_accretion} ($\teff\simeq2950$~K), the models predict
a depth of the TiO feature corresponding to [$770~index$]$=-0.8$~mag, and
by adding $10\%$ of accretion luminosity one obtains a shift of the index
of $\sim0.06$~mag. In fact, from Figure \ref{fig:index_teff} and Equation
\ref{equation:teff_vs_index}, [$770~index$]$=-0.8$~mag corresponds to a lower
$\teff$, about 2700~K.
Therefore we can associate this $\teff$ with a shift of $\sim0.06$~mag.
Then, we readjust
the ratio $L_{\rm accr}/ L_{\rm star}$, using the updated $\teff$,
according to $L_{\rm star}\propto\teff^4$.

Finally, we can recast the result from a more observational perspective,
in which we measure [$770~index$] for a given star and we do not know
the true value of $L_{\rm accr}$ for that star.
Deriving $\teff$ via Equation \ref{equation:teff_vs_index},
which is defined for $L_{\rm accr}=0$,
we will overestimate the true $\teff$ because accretion will cause
the TiO feature to be filled-in and consequently we will measure a
higher [$770-index$].  Figure
\ref{fig:accretion_bias} shows the resulting systematic
overestimation of the derived $\teff$ for various $L_{\rm accr}$.
We find again that the
lower the stellar $\teff$, the larger the error in the $\teff$ estimate due
to accretion. However, if one excludes very strong accretors ($L_{\rm accr}
\simeq L_{\rm star}$), the error in $\teff$ is very small. For instance,
for $L_{\rm accr} = 10\% L_{\rm star}$, the $\teff$ error is always smaller
than half spectral subtype, and always $<40$~K above the H-burning limit
($\teff \sim 3000$~K).
The  $\teff$ difference associated to an
error of 1 spectral type was computed as the difference between the temperature of two contiguous spectral subtypes.
Similarly, the $\teff$ differences corresponding to
 $1/2$ and $1/4$ spectral subtypes (dotted lines in Figure
\ref{fig:accretion_bias}) is equal to that for 1 spectral type, scaled down by a factor of 2 and 4.


In conclusion, the presence of optical excess due to
accretion has a negligible effect on the stellar parameters derived from
our medium-band survey. Thus, we will neglect accretion in the following
analysis, and proceed to derive $\teff$ based on the [$770~index$] from
Equation \ref{equation:teff_vs_index}.

\subsection{Accuracy of $\teff$ from previous works}
Despite having shown that accretion excess not not alter significantly
the measured [$770~index$], Figure \ref{fig:index_teff}
showed that accreting ONC sources are displaced in the
$\teff$ vs. [$770~index$] plane.
The displacement is such that the accretors lie above the non-accretors,
meaning that the spectroscopic $\teff$
plotted on the y-axis of Figure \ref{fig:index_teff}
have been on average overestimated for accreting ONC sources.

This overestimation of spectroscopically determined $\teff$ for
accreting sources is likely the result of incomplete removal of the
optical veiling in the spectra of these sources. This
excess, that we demonstrated above to be minimized at 7700\AA  (although it may still be present in extreme cases, see, e.g., \citealt{fischer2011}), is likely larger at bluer wavelengths, where other
molecular features are typically used to measure the spectral types of
M-type stars.  This veiling, if not removed or if incompletely removed,
will fill in the molecular
absorption lines and lead to a systematic overestimation of the spectral types.

By utilizing the [$770~index$] to derive $\teff$ we are instead largely
immune to these effects, and we are now able to correct the previous
spectroscopically derived $\teff$ via equation Equation
\ref{equation:teff_vs_index}.

In conclusion, our technique to measure stellar $\teff$ is able to derive a more accurate estimate of $\teff$ than optical spectroscopy, being less biased by a possible veiling excess.

\subsection{Calibration of the [$A_V~index$] and derivation of extinctions}
\label{subsection:Av_calibration}

\begin{figure*}
\includegraphics[width=0.75\columnwidth,angle=90]{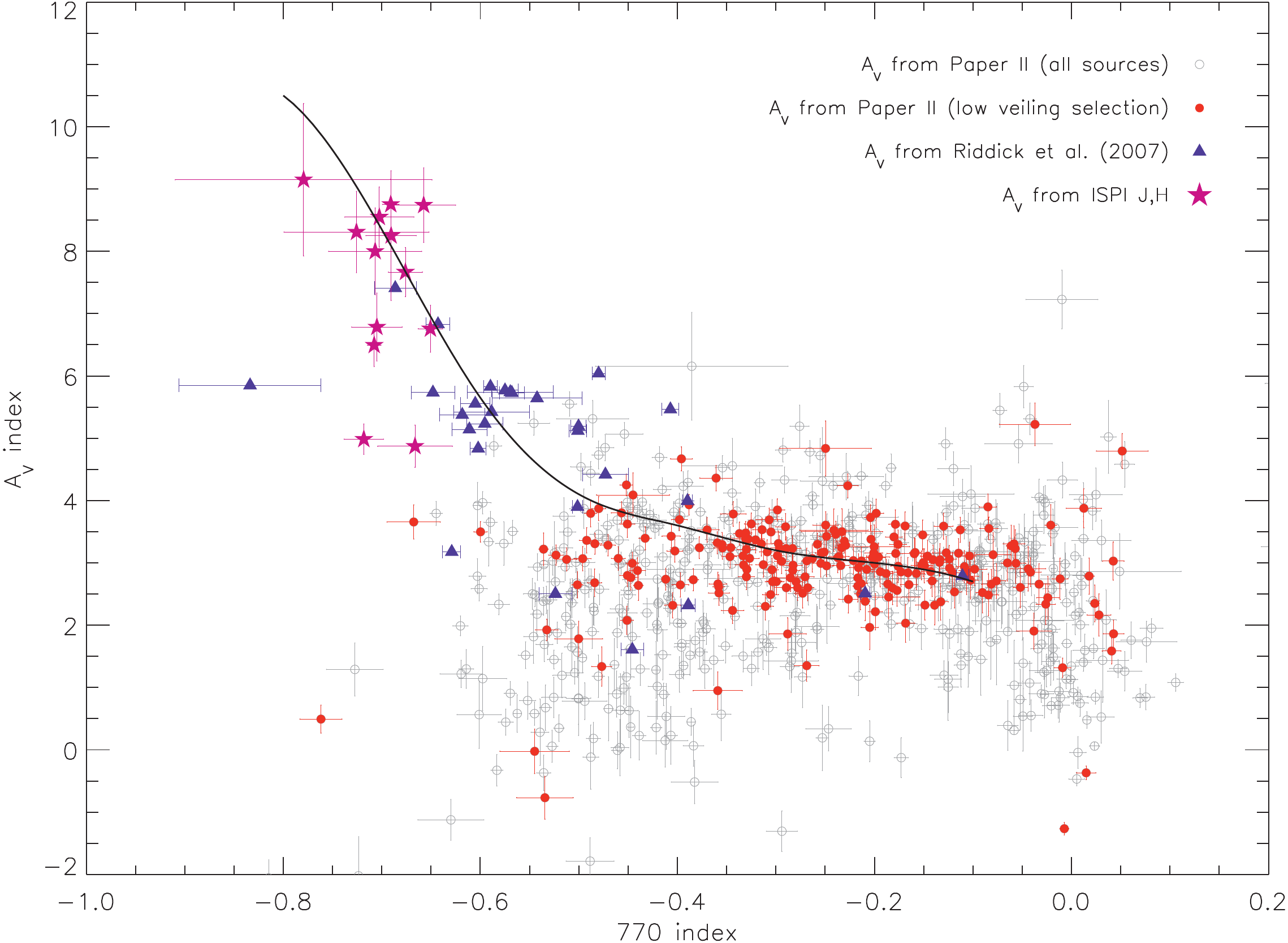}
\includegraphics[width=0.75\columnwidth,angle=90]{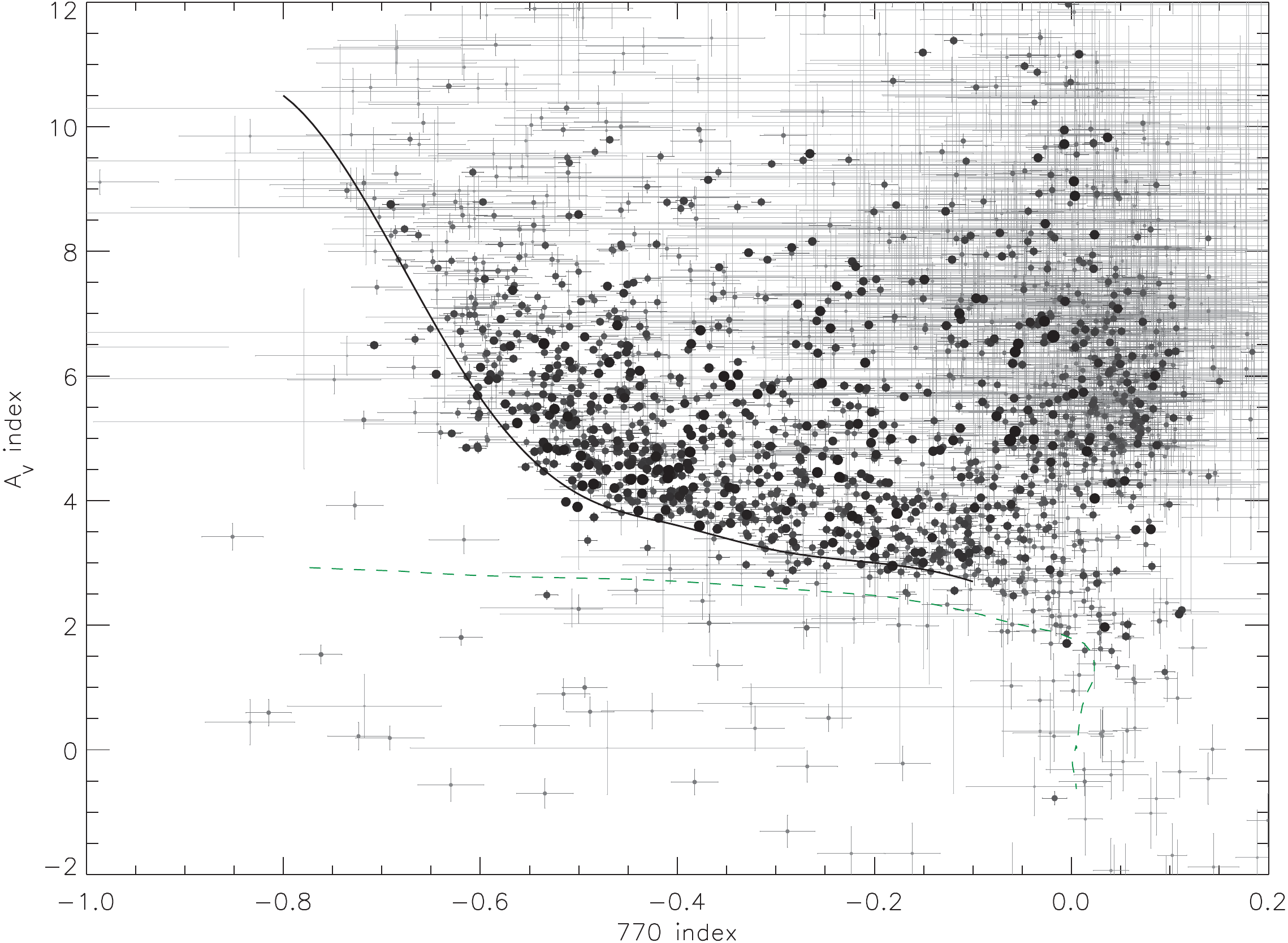}
\caption{The color-color plot expressed in units of $[770~index]$ (reddening independent) and $[A_V~index]$ (parallel to the reddening direction). \emph{Left panel:} sample of extinction-corrected sources; $A_V$ values are from \citet{riddick2007}, the $JHK$ photometry of \citet{robberto2010}, and the optical analysis of Paper~II. For the latter, accretors and candidate non accretors are distinguished as in Figure \ref{fig:index_teff}. The solid line represents our best fit of the data, i.e. the empirical isochrone for $A_V=0$. \emph{Right panel}: same diagram, reporting all the sources in our catalog, with no correction for extinction. These are color coded according to their photometric errors. The entire population, with the exception of very few sparse sources, is located above the isochrone, at positive values of extinction.     \label{fig:Aindex_calib}}
\end{figure*}

\begin{figure}
\epsscale{1.1}
\plotone{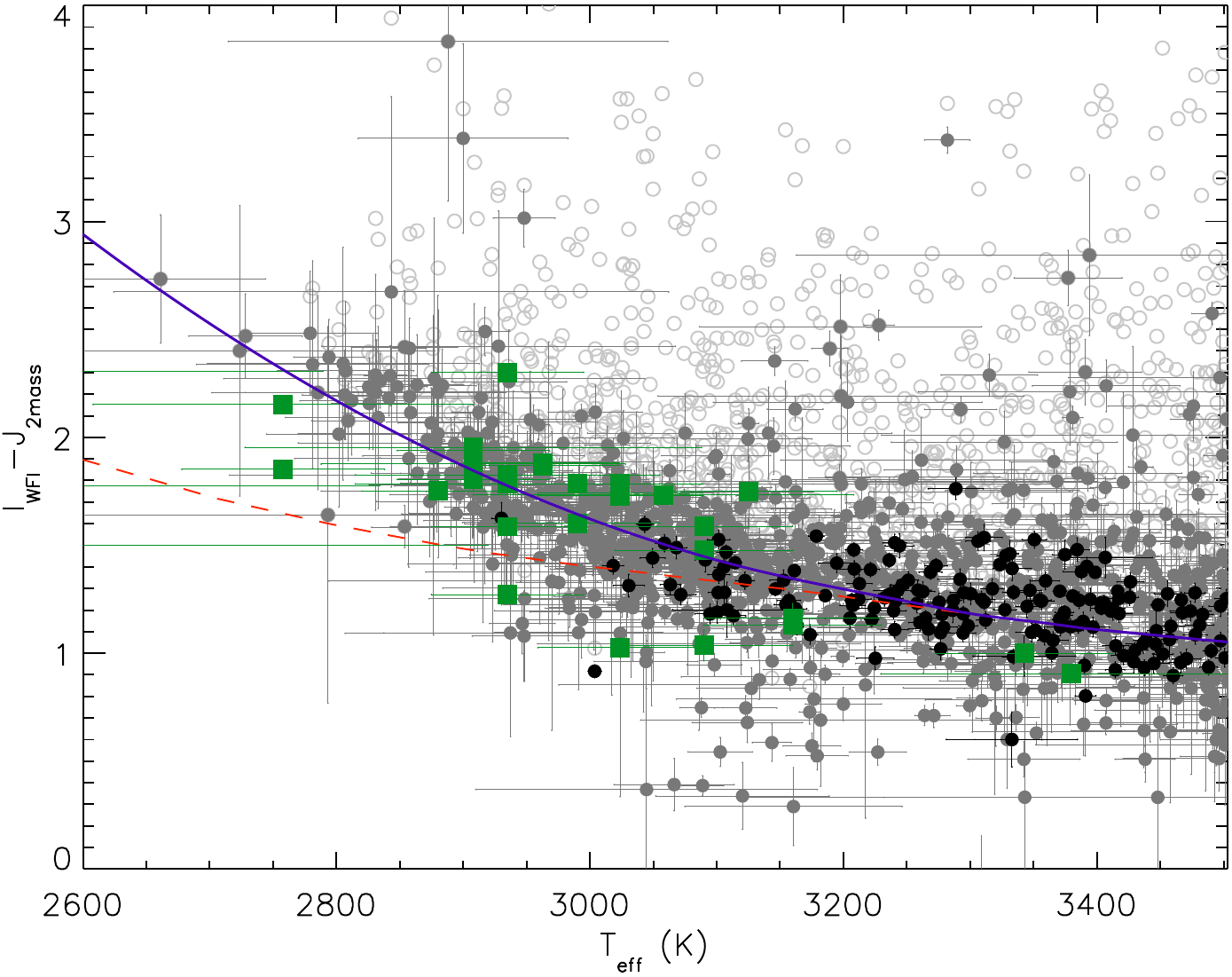}
\caption{$(I-J)$ vs $\teff$: the open dots are the measured colors, the filled gray dots are the de-reddened ones. Black dots are dereddened colors for sources with know estimate of accretion excess, showing no significant excess. Green squares are sources whose $\teff$ is from \citet{riddick2007}. The dashed line is the prediction from synthetic photometry on the BT-Settle atmosphere models of \citet{allard2010}; the solid line is the empirical fit to the data.  \label{fig:ij_vs_teff}}
\end{figure}

In section \ref{subsection:teff_calibration} we derived an empirical
relation between the $[770~index]$ and the stellar $\teff$, and in section \ref{subsection:accretion} we explored the effects of accretion luminosity on there derivations. Here,
similarly, we find the relation between the $[A_V~index]$, obtained in
Equation \ref{equation:Avindex}, and the actual extinction $A_V$ of the
ONC stars, in magnitudes.
In the color-color diagram, the extinction $A_V$ of a star
is the distance, along the reddening direction, between the observed
position and the isochrone, whereas the [$A_V~index$] is the distance, also
along the same direction and scaled by the appropriate $A_\lambda/A_V$, between the observed colors and the abscissa
$(770-753)=0$. Therefore the [$A_V-index$] and the true $A_V$ differ only
by a constant, which depends on the [$770~index$] (i.e., it is
a function of $\teff$).

To this end, we plot the extinction-corrected ONC stars with independent $A_V$ estimates
from Paper~II and from \citet{riddick2007}
in the [$A_V~index$] vs. [$770~index$] plane
(Figure \ref{fig:Aindex_calib}, left-hand panel).
We also use the ISPI $JHK$ photometry of the ONC presented
in \citet{robberto2010}, from which we use stars with the lowest $\teff$
(low $[770~index]$)
and with the smallest photometric errors.
For the coolest $\teff$ considered here, the $J$ vs $(J-H)$ isochrone
from \citet{robberto2010} is
vertical, and its color ($J-H\simeq0.65$) is independent of age,
therefore $A_V$ can be derived by simply dereddening the sources
on the isochrone. By not including the $K$-band photometry from
\citet{robberto2010},
the infrared-derived extinctions should not be strongly affected by the
possible presence of disk excess.
The thick line in the figure is our fit to
the data, i.e., the empirical isochrone corresponding to $A_V$=0. This also
represents the offset to be subtracted from the measured $[A_V~index]$,
as a function of the [$770~index$], to compute the true $A_V$.

Among the points in Figure \ref{fig:Aindex_calib}
from our previous Paper~II, we can distinguish accreting sources from
non-accretors, just as in Figure \ref{fig:index_teff}. In Section
\ref{subsection:teff_calibration} we suggested that previous
spectroscopically derived $\teff$
estimates are systematically overestimated for accreting members in the
M-type range. This is now confirmed by Figure \ref{fig:Aindex_calib}: whereas
the extinction-corrected non-accretors lie on a curve consistent with the
lower edge of the non extinction-corrected population (right panel of the
same Figure), for accretors we find $A_V$ estimates systematically
overestimated. This is because if $\teff$ is overestimated, the intrinsic
colors are underestimated, and therefore the color excess attributed to
extinction is overestimated leading to higher $A_V$.

Figure \ref{fig:Aindex_calib}, right-hand panel, shows again the
empirically calibrated isochrone (i.e., for $A_V=0$),
together with all the stars in our catalog. Unlike in the
left-hand panel, we have not subtracted the extinction here.
Overall the
population lies above the line, i.e., at positive values of $A_V$.
As would be expected if the isochrone is correct, the
distribution of sources above the isochrone also appears uniform for
different $[770~index]$.  There is an overabundance of highly
reddened stars at the highest useful limit of the index ($[770~index]
=-0.1$), i.e., at the highest limit of the $\teff$ range accessible with
our medium-band photometry. This, as will be shown below,
is a residual late-type background contaminant population.

\section{The H-R diagram}
\label{section:HRD}
\begin{figure*}
\epsscale{0.75}
\plotone{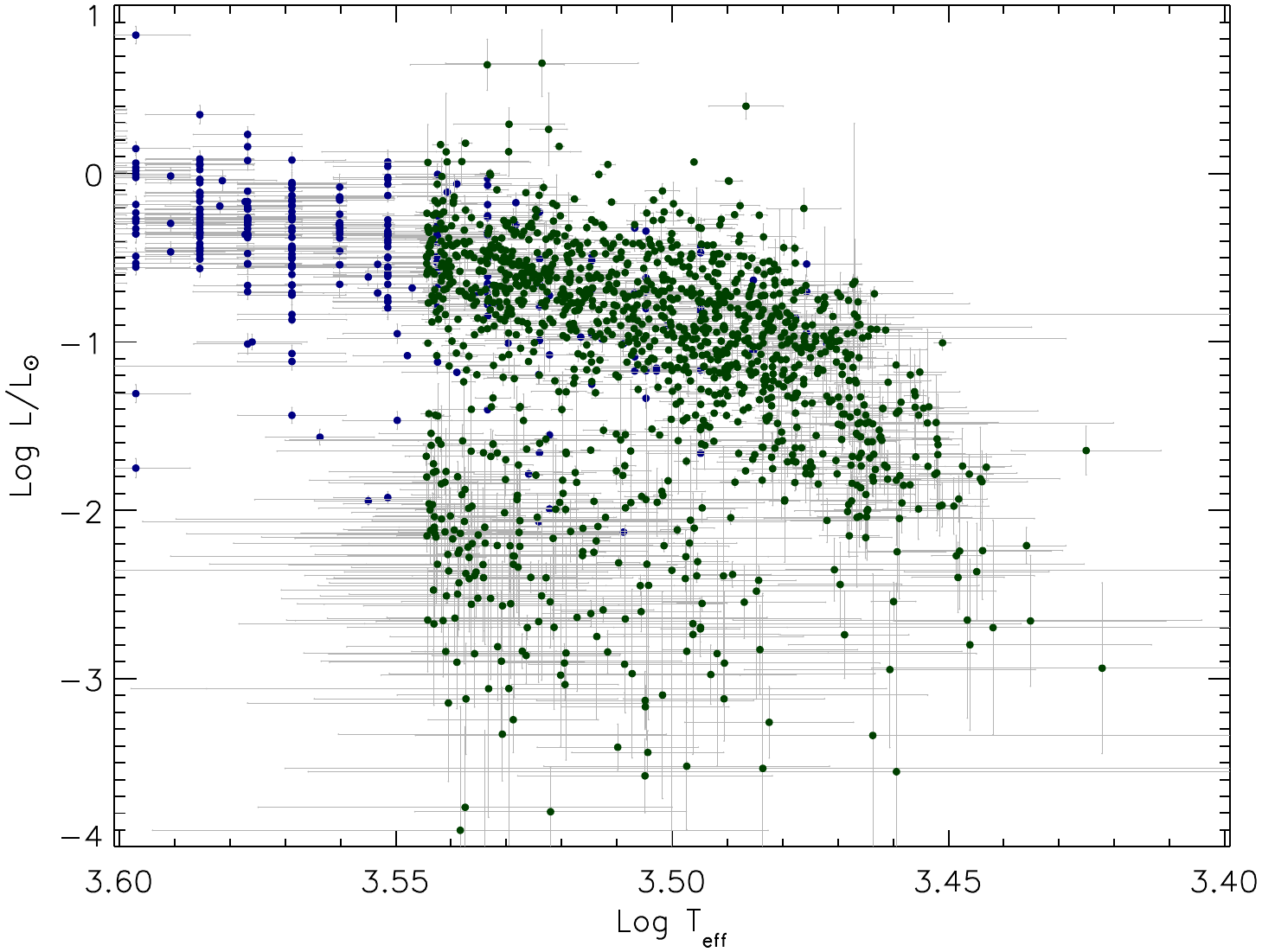}
\caption{The new H-R diagram of the ONC. Green dots are the new stars whose stellar parameters have been derived in this work, blue dots are members from Paper~II, mostly having a large enough $\teff$ to not be derivable using the $[770~index]$, or saturated in our new WFI photometry.\label{fig:HRD}}
\epsscale{1}
\end{figure*}

\begin{figure}
\epsscale{1.1}
\plotone{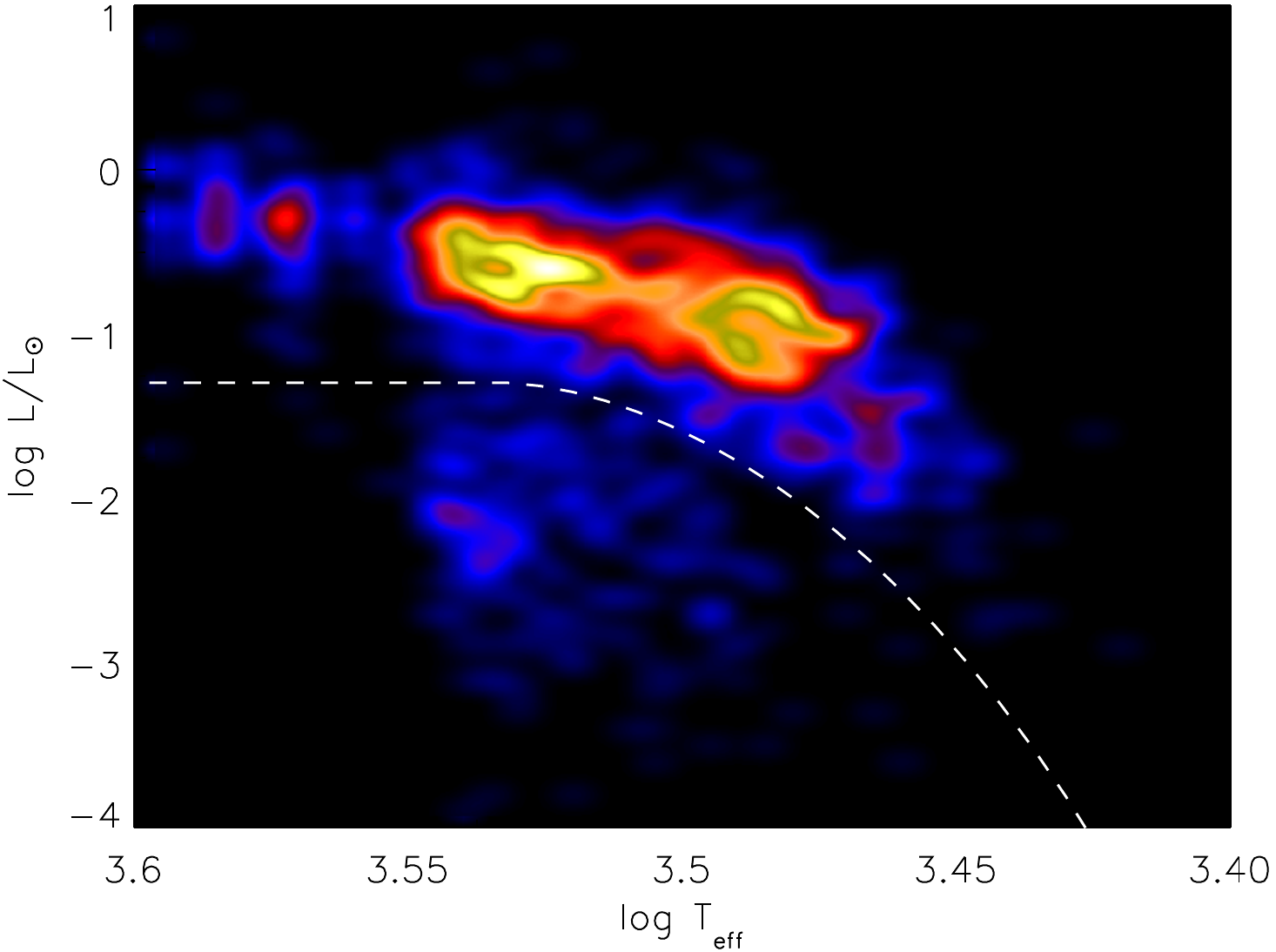}
\caption{Stellar density map in the H-R diagram. The dashed line marks our adopted separation between bona fide members and contaminants.  \label{fig:contaminants_selection}}
\end{figure}

\begin{figure}
\epsscale{1.1}
\plotone{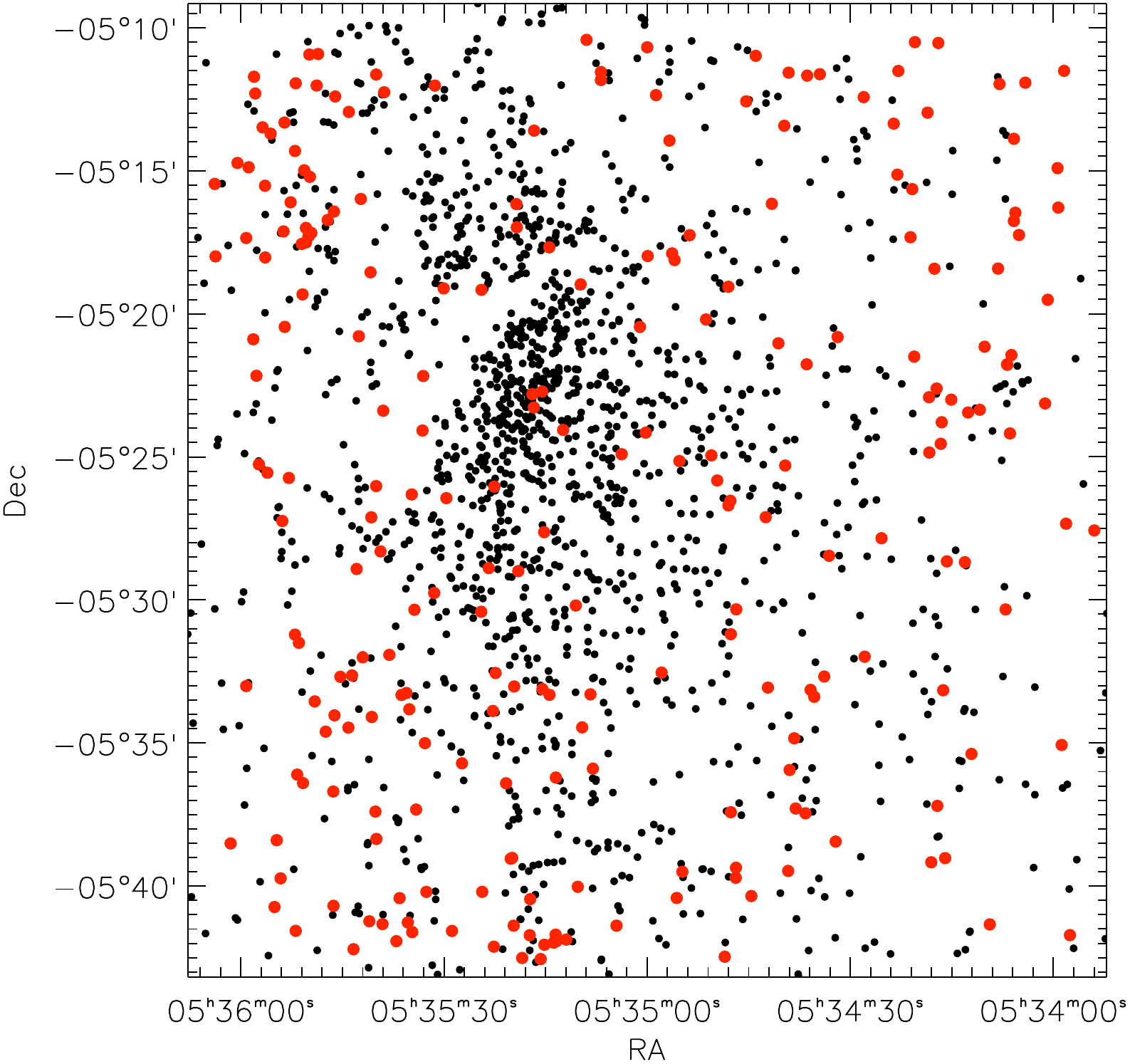}
\caption{Spatial distribution of the ONC members placed in the H-R diagram (black dots). The red circles mark the position of the faint sources identified as background sources.  \label{fig:contaminants_radec}}
\end{figure}

Having derived two empirical relations, one to obtain the $\teff$ from the
$[770~index]$ for stars in the spectral subtype range M3--M8.5, the other
to derive $A_V$ once the $[770~index]$ is known, we now derive the stellar
parameters for the sources in our catalog.

Sources lying below the isochrone of
Figure \ref{fig:Aindex_calib}b appear to have negative extinctions, which is
unphysical; 93 stars showing an extinction in the range $-2$~mag$<A_V<$0~mag
are therefore assigned $A_V=0$,
essentially shifting the measured
colors to the closest physically meaningful solution. Another few stars
with $A_V<-2$~mag were removed from the sample, being likely spurious
detections or faint sources with very inaccurate photometry.
The fraction of sources showing negative extinction
(about 5\%) is much smaller than in previous works (e.g., Paper~II or
\citealt{hillenbrand97}), validating the improved accuracy of our methods
to derive stellar parameters. In total, we derived $\teff$ and $A_V$
for 1280 sources, whereas only 544 of them had a previously assigned $\teff$.
Merging the new $\teff$ and $A_V$ with those of our
previous work, our sample counts 1807 sources with both $\teff$ and $A_V$
from which we can derive an updated H-R diagram for the ONC extending to
well below the H-burning limit.

\subsection{Bolometric corrections}
\label{subsection:bolometric_corrections}

In order to convert the de-reddened magnitudes into $\log L/L_{\odot}$,
knowledge of the bolometric corrections (BCs) as a function of $\teff$
is required. Unfortunately, we cannot rely for such relations on previous
works \citep[e.g.,][]{flower96,bessel98} for several reasons: 1) the
photometric bands used here, including $I-$band,
are non-standard; 2) optical BCs usually do not extend in the
BD temperature range; and 3) if the optical stellar colors are age dependent,
as suggested in Paper~II, the BCs may be also. Thus, BCs valid for
main-sequence dwarfs
might not be adequate for a young populations such as the ONC. Instead,
we use synthetic photometry to derive our BCs, according to the formula:
\begin{equation}
\label{equation:BC}
{\rm BC} = 2.5\cdot \log\bigg[\frac{\int f_\lambda (T)\cdot S_\lambda d \lambda}{\int f_\lambda (T) \cdot d \lambda}\bigg/ \frac{\int f_\lambda (\odot)\cdot S_\lambda d \lambda}{\int f_\lambda (\odot)\cdot d \lambda}\bigg],
\end{equation}
\noindent where $f_\lambda (T)$ are synthetic spectra, and $S_\lambda$
the filter throughput.

As we have shown above (see also the discussion in
Paper~II), current atmosphere models do not correctly predict
the optical broad band colors for cool stars. This clearly affects the
computation of BCs, in particular BCs in $I-$band, $BC_I$.
On the other hand, the bolometric flux for a given
synthetic spectrum will be trivially correct at a given $\teff$ since,
by definition, $L\propto R^2 T^4$. The radii $R$, which could also be
inaccurate from the evolutionary models, do not affect the BCs since they
introduce an identical proportionality factor in both the numerator and
denominator of equation \ref{equation:BC}.

From the comparison of our data with the synthetic isochrones (see Figure
\ref{fig:color_color_models}) we cannot ascertain the correctness of
the synthetic $I$-band fluxes. It is true that the isochrones are
unable to reproduce the data in the color-color diagram, but the
shortcomings may be principally in our medium bands and not in the
$I$-band. At the same time, the
BT-Settle models \citep{allard2010} have been validated in
the NIR for dwarf stars, such that the predicted
$J$ and $K$ fluxes match the observations down to BD
masses. Thus, if the predicted $I-$band magnitudes are also correct,
we would expect the predicted $(I-J)$ color to be also
consistent with the observations across all $\teff$.

We perform this test using the $J$-band photometry of the ONC
from \citet{robberto2010}, together with our $I-$band magnitudes
and (as in Section \ref{subsection:Av_calibration}) the $A_V$
estimates from previous works. Figure \ref{fig:ij_vs_teff} shows the
dereddened $(I-J)$ color as a function of $\teff$ for our ONC sample, in
comparison with the same quantity computed using the BT-Settle atmospheres,
assuming a \citet{bcah98} 1~Myr isochrone to constrain the surface gravity
as a function of $\teff$.  As in Figure \ref{fig:ij_vs_teff}, we highlight
sources showing no or little accretion excess, except at the coolest $\teff$
($\teff \lesssim 3000$~K), for which we do not have estimates of accretion
from Paper~II.  For $\teff \gtrsim 3200$~K, the isochrone fits
the data well. However, the match is worse at lower $\teff$,
the isochrone predicting lower $(I-J)$ than shown by our data.
We checked that this conclusion is not strongly dependent on $\log g$.

Trusting the predicted $J-$band fluxes, as explained above, we conclude that
the BT-Settle atmosphere models tend to systematically overestimate the
$I-$band fluxes, and this effect is larger for cooler $\teff$. In
Figure \ref{fig:ij_vs_teff} we also show the empirical fit to the data. The
difference between the two lines represents the empirical correction needed
to calibrate the predicted $I$-band synthetic photometry using the BT-Settle
models. Whereas this is small for $\teff \gtrsim 3000$~K,
in the BD $\teff$ range the correction increases up to 1
magnitude. Thus, we correct the computed BCs, derived from Equation
\ref{equation:BC}, applying this offset as a function of $\teff$.

Finally, we derive the bolometric luminosities for all our sources from:
\begin{equation}
\label{equation:HRD}
\log L/L_\odot = 0.4\cdot(M_{I,\odot}-I+A_I+DM+BC_I(T_{\rm eff}))
\end{equation}
\noindent where $M_{I,\odot}=4.03$ is the absolute magnitude, in WFI $I$
band, of the Sun (Paper~II) and DM=8.085 is the distance modulus, adopting
$d=414\pm7$~pc from \citet{menten2007}.

The resulting H-R diagram is shown
in Figure \ref{fig:HRD}, where we show separately the stars whose
parameters have been derived from the new data presented in this work,
and those, generally with $T_{\rm eff}>3500$~K, whose $\teff$ and
$L$ are from Paper~II. The apparent discontinuity in the number of
sources around $\log\teff\approx 3.54$
is mainly due to the much higher completeness in the number of low-mass
stars with $\teff$ estimates provided by the present work
(see Section \ref{subsection:completenessHRD}).

\subsection{Background contamination}
\begin{figure*}
\epsscale{0.8}
\plotone{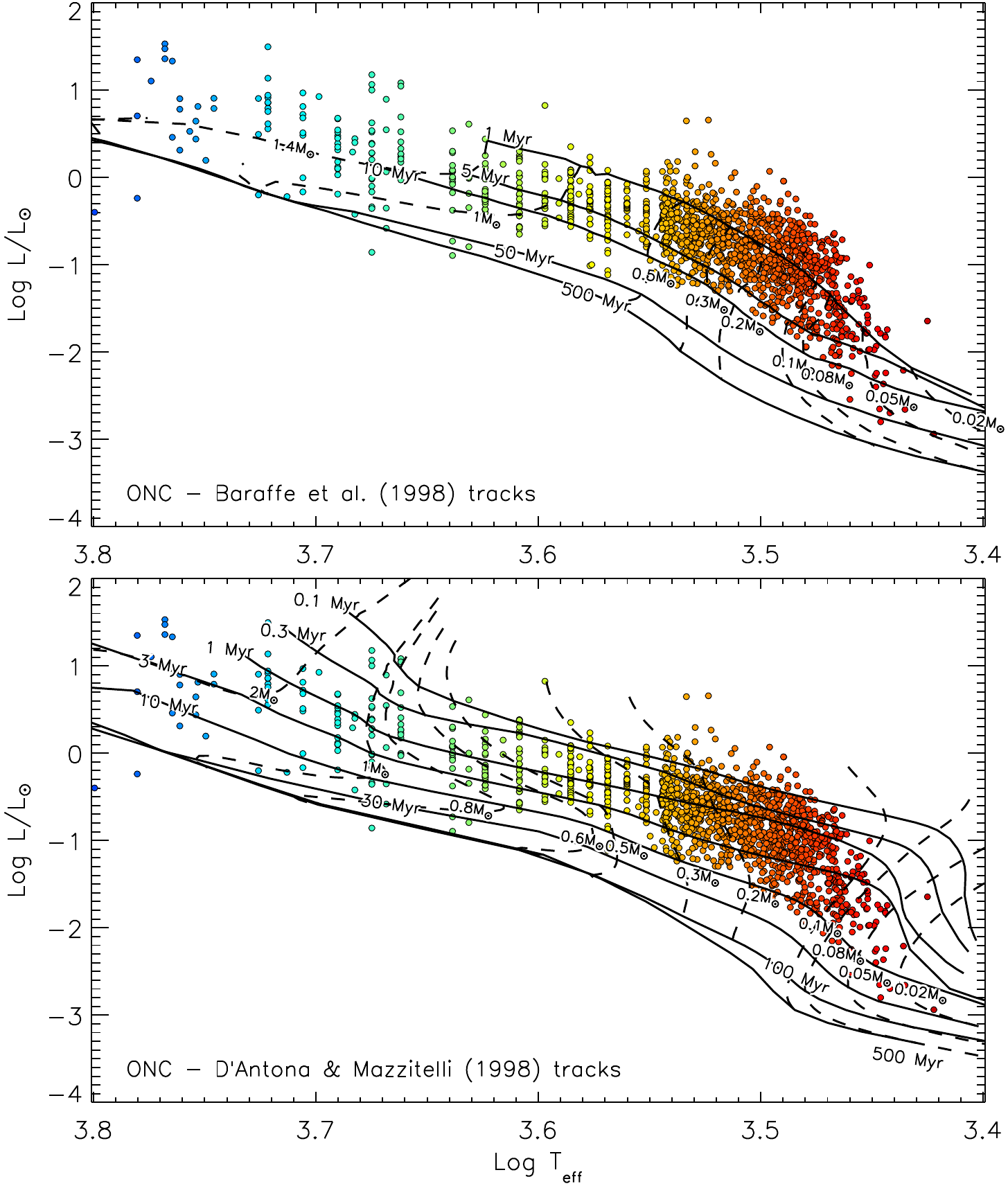}
\caption{The H-R diagram of the ONC members, after the removal of contamination from background stars. Evolutionary models are from \citet{bcah98} (upper panel) and \citet{dantona98} (lower panel). \label{fig:HRD_with_models}}
\epsscale{1}
\end{figure*}

Despite the large column density of the Orion molecular
cloud, which produces an extinction up to $A_V=100$~mag
\citep{Bergin1996,scandariato_extinction}, contamination from background
stellar populations starts to become non negligible at faint magnitudes in the ONC. In fact, as suggested by previous works,
\citep[e.g.][]{robberto2010,hillenbrand-carpenter2000,muench2002}, the
fraction of background contaminants versus members increases towards the substellar mass range in the ONC. This is both due to the increase in the number
of faint background sources seen through the Orion molecular cloud, and
the decrease of actual members of the young cluster in the BD mass range,
because of the declining initial mass function (IMF) at substellar masses
(see below).

The contamination from background sources is evident as a large
number ($\gtrsim$200 stars) of faint sources in our H-R diagram (Figure
\ref{fig:HRD}), which are well separated from the cluster sequence. Given the
relatively low $\teff$ for these objects ($\sim 3500$~K$<\teff<\sim2900$~K),
they could be either cool dwarf stars or red giants, the latter more easily detectable at large distances and through the Orion Molecular Cloud. The relative number of these sources, relative
to young cluster members, becomes dominant at our detection limit.

A small number of these relatively less-luminous sources
might be actual ONC members, whose flux may be masked
(e.g., by edge-on circumstellar disks), and therefore are visible
only in scattered light. This phenomenon has been observed in
several young star-forming regions \citep[see e.g.,][and references
therein]{guarcello2010,kraus-hillenbrand2009}; however the expected fraction of such objects is very small, including no more than a few percent of the population \citep{demarchi2011}.
On the other hand, the faint sources we detect below
the young sequence are about 16\% of the total number of stars with
$\teff<3200$~K. Therefore, we can assume that most of these
sources are actually contaminants. Moreover, removing a small fraction
of faint members erroneously considered as background contaminants will
not affect significantly the IMF derived
in this work, given the small expected number of these objects compared
to the entire studied population.

We proceed to attempt to remove
the background contamination from the H-R diagram. To this purpose, we
define a cut in the H-R diagram
such that stars satisfying the following criterion
are considered as non-members:
\begin{eqnarray*}
  \log L <  -1.93\ & & \ \ \log\teff>3.54  \nonumber \\
  \log L <  -5.607&  &  \nonumber \\
   +65.848& \cdot(\log\teff-3.4) &   \nonumber \\
   -253.02& \cdot(\log\teff-3.4)^2&   \ \ \log\teff<3.54
\end{eqnarray*}
\noindent This is shown in Figure \ref{fig:contaminants_selection}. All
sources located below this line are regarded as non-members and excluded from
further analysis. Figure \ref{fig:contaminants_radec} presents the spatial
distribution of contaminants in comparison with that of the candidate
members. Whereas the latter are centrally concentrated, with an evident
north-south elongation, contaminants appear uniformly distributed. Also,
we find a slight overabundance of contaminants on the east side of our
field of view. The reason for this is that the extinction provided
by the Orion Nebula is higher on a central north-south strip, centered on
the ONC \citep{scandariato_extinction,Bergin1996}. Therefore the density
of background objects is expected to be higher on the two sides of this
band. Here, the nebular emission of the OMC is significantly brighter on
the west side of our FOV, and this decreases the detection limit in this
area. Therefore the east edge of our FOV is expected to show the most
prominent density of background contaminants.

In Figure \ref{fig:HRD_with_models} we present once again the H-R
diagram of the ONC, now including only our candidate members, together with
PMS evolutionary models. In particular, we use isochrones and tracks from
either \citet{bcah98} or \citet{dantona98}. These models, unlike e.g.,
the \citet{siess2000} or the \citet{pallastahler99}, also extend into
the substellar mass range, and therefore are more suitable for our
sample, which reaches masses as low as 0.02~M$_\odot$ (20M$_{Jup}$).

\subsection{Completeness in the H-R Diagram}
\label{subsection:completenessHRD}
\begin{figure*}
\plottwo{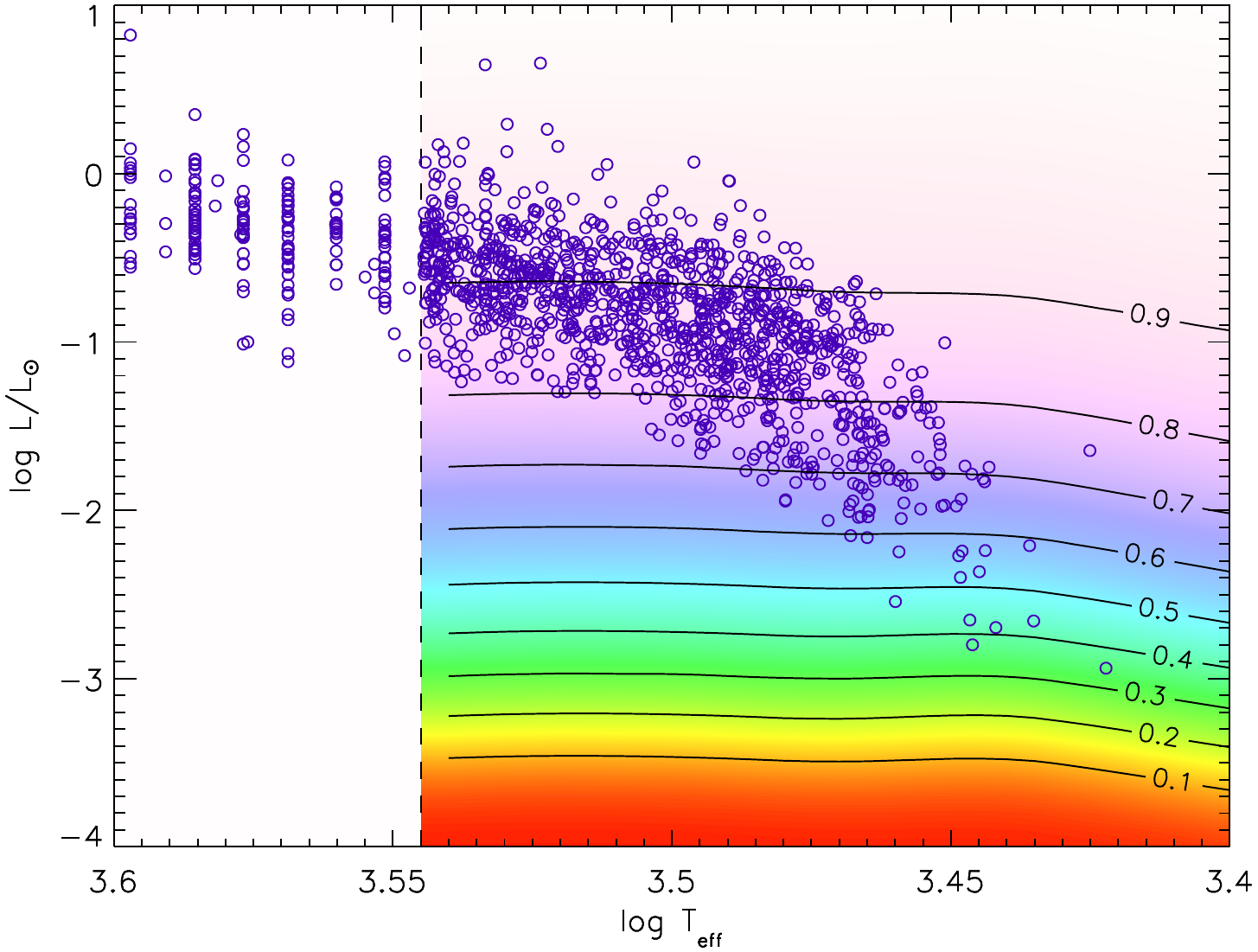}{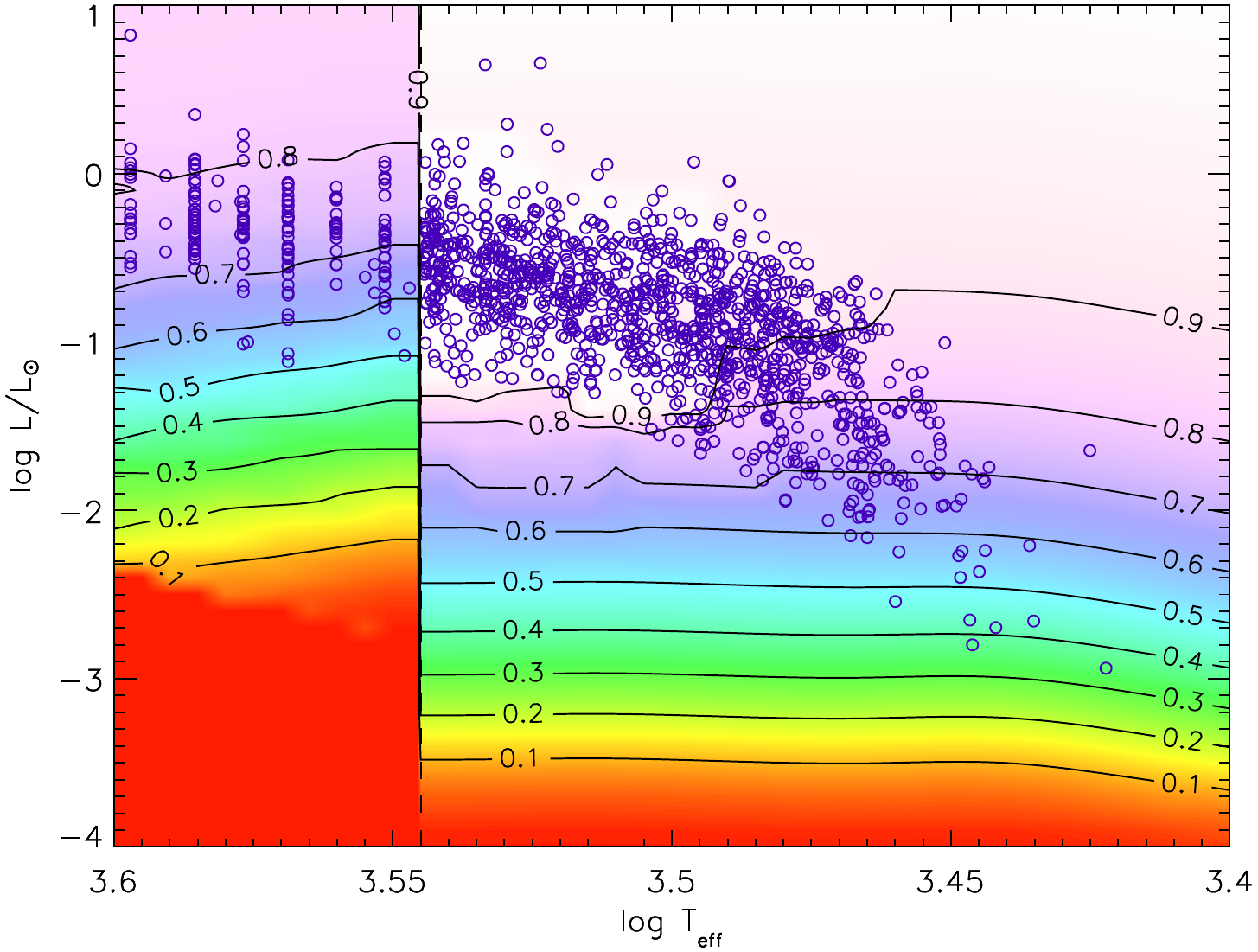}
\caption{Completeness function in the H-R diagram, computed accounting for photometric detection and differential reddening as described in the text. The left panel shows the result as computed with our monte-carlo simulation, valid for the spectral types obtained from the medium-band photometry analyzed in this work, for $\log\teff<3.54$ ($\teff < 3500$~K). The right panel shows the correction of this completeness after accounting for the additional stars placed in the HRD whose stellar parameters are taken from Paper II.   \label{fig:completeness_hrd}}
\end{figure*}

In Section \ref{subsection:photometric_completeness} we have derived the
photometric completeness of our survey. Starting from this result, we now
evaluate the completeness function in the H-R diagram. To do this,
we transform the H-R diagram (i.e., $\teff$ and $\log L$) into our
observational space (i.e., the 753, 770, and $I$-band magnitudes), and
assess the extent to which reddening causes sources of various
intrinsic $\teff$ and $\log L$ to be missed in our survey.

We set a uniform, dense grid of $\log \teff$ and $\log L$ bins in the
H-R diagram, limited to
the region $\log \teff<3.54$. For each bin, we apply in reverse
the relations derived in Section \ref{subsection:teff_calibration}
and \ref{subsection:Av_calibration} to derive from $\teff$ and $\log L$
the $[770~index]$, the $[A_V~index]$, and (using the BC) the
unreddened $I$-band magnitude, $I_0$. Then, by
inverting equation \ref{equation:770index} and \ref{equation:Avindex},
we also derive the magnitudes $m_{753}$ and $m_{770}$. Thus we have
associated to each pair of H-R diagram parameters ($\log \teff$,$\log L$) the
triplet of intrinsic magnitudes ($m_{753},$$m_{770}$,$I_0$) for $A_V=0$.

Next we apply reddening to these magnitudes using
the empirical reddening distribution
of the ONC stellar population. As in Paper~II, we limit ourselves to the
most luminous half of the ONC population,
as these stars sample the full range of $A_V$.
With a Monte Carlo approach, we draw 1000
$A_V$ values from the reddening distribution and apply them to
the intrinsic magnitudes ($m_{753},$$m_{770}$,$I_0$) associated
to each point of the H-R diagram. For every $i$-th ($1<i<1000$) triplet of
reddened magnitudes, we compute the photometric completeness $C_{753}(i)$,
$C_{770}(i)$, $C_{I}(i)$. The product of these three, $C(i)$, averaged over
the 1000 values of $A_V$, provides the ``differential reddening normalized''
completeness for that particular bin in the H-R diagram.

The result is
shown in Figure \ref{fig:completeness_hrd}a. We find that the completeness
is mainly dependent on $\log L$, while its dependence on $\teff$
for a given luminosity is small and appears only at low temperatures
($\log\teff\lesssim3.45$). We note also that the completeness is rather
high ($\sim50\%$) at the low-mass end of our catalog, and thus even
without correcting for the completeness, the observed number of ONC
members in the substellar mass range is correct to within a factor of $\sim$2.

Our H-R diagram also includes sources whose stellar parameters are taken
from Paper~II instead of our new medium-band photometry. For all members
with $\log \teff>3.54$, we simply adopt the completeness function in
the HRD derived in that paper. In addition, there is a group of stars
with $\log\teff<3.54$ but which are not present in
our new medium-band photometry and thus
whose stellar parameters are from Paper~II.
The presence of these sources
increases the overall completeness in this region of the HRD relative
to that computed as above. To account
for this, we divide the HRD into a grid with spacing 0.01~dex in $\log \teff$
and 0.1~dex in $\log L$, and we count in each bin the relative number of
sources from Paper~II versus members classified from our new data. This
ratio, smoothed in $\log \teff$ and $\log L$, is added to the completeness
shown in Figure \ref{fig:completeness_hrd}a.

The final result is shown in Figure
\ref{fig:completeness_hrd}b. The evident discontinuity in completeness at
$\log\teff=3.54$ is due to the fact that for larger $\teff$ the stellar
parameters are obtained from optical spectroscopy on an
incomplete fraction of sources.
From the two-dimensional completeness shown in Figure
\ref{fig:completeness_hrd}b, we are able to assign a completeness
correction to each our sources, allowing us finally to derive a
completeness corrected initial mass function for the ONC.

\section{The initial mass function}
\label{section:IMF}

\begin{figure*}
\plottwo{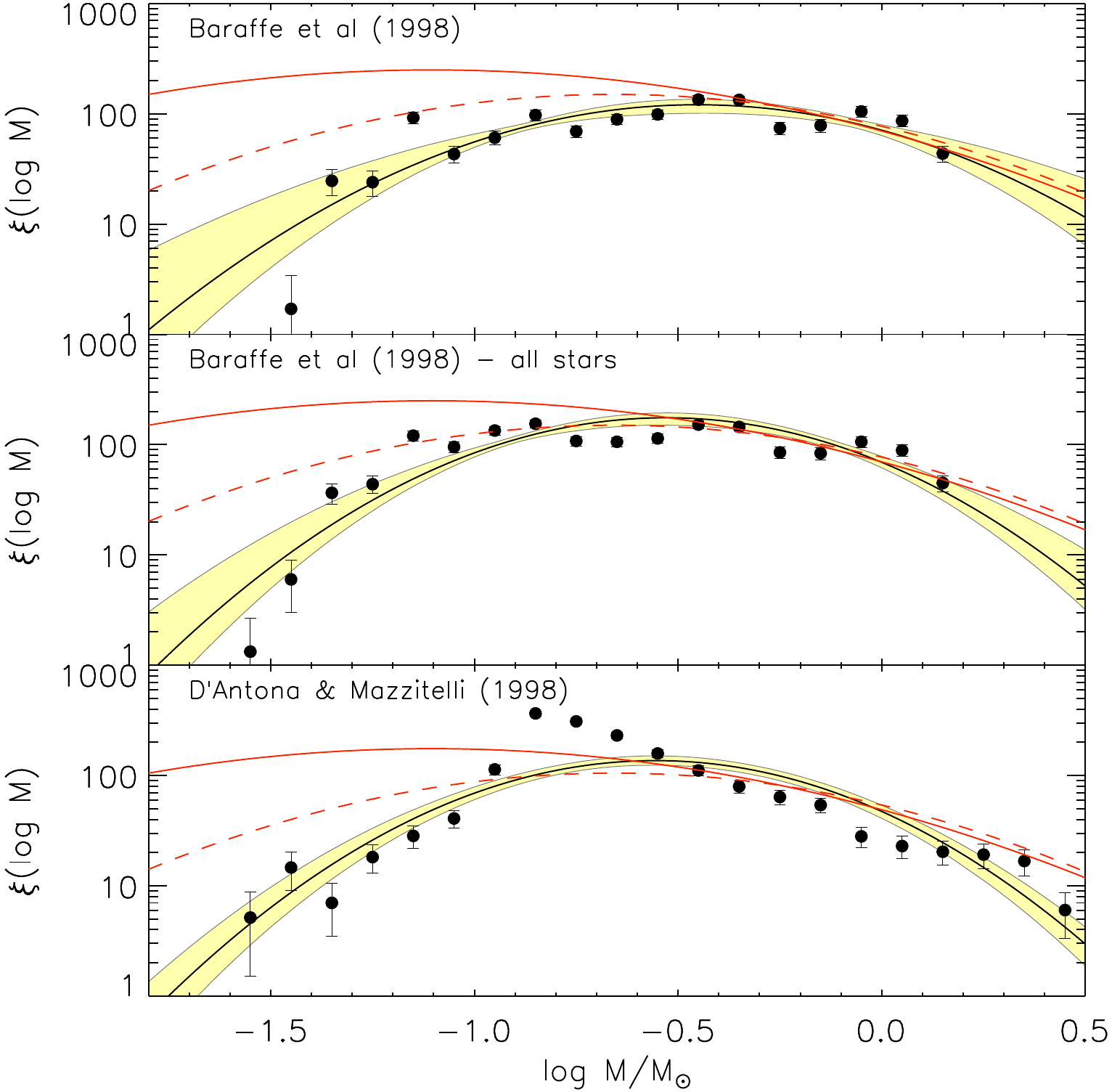}{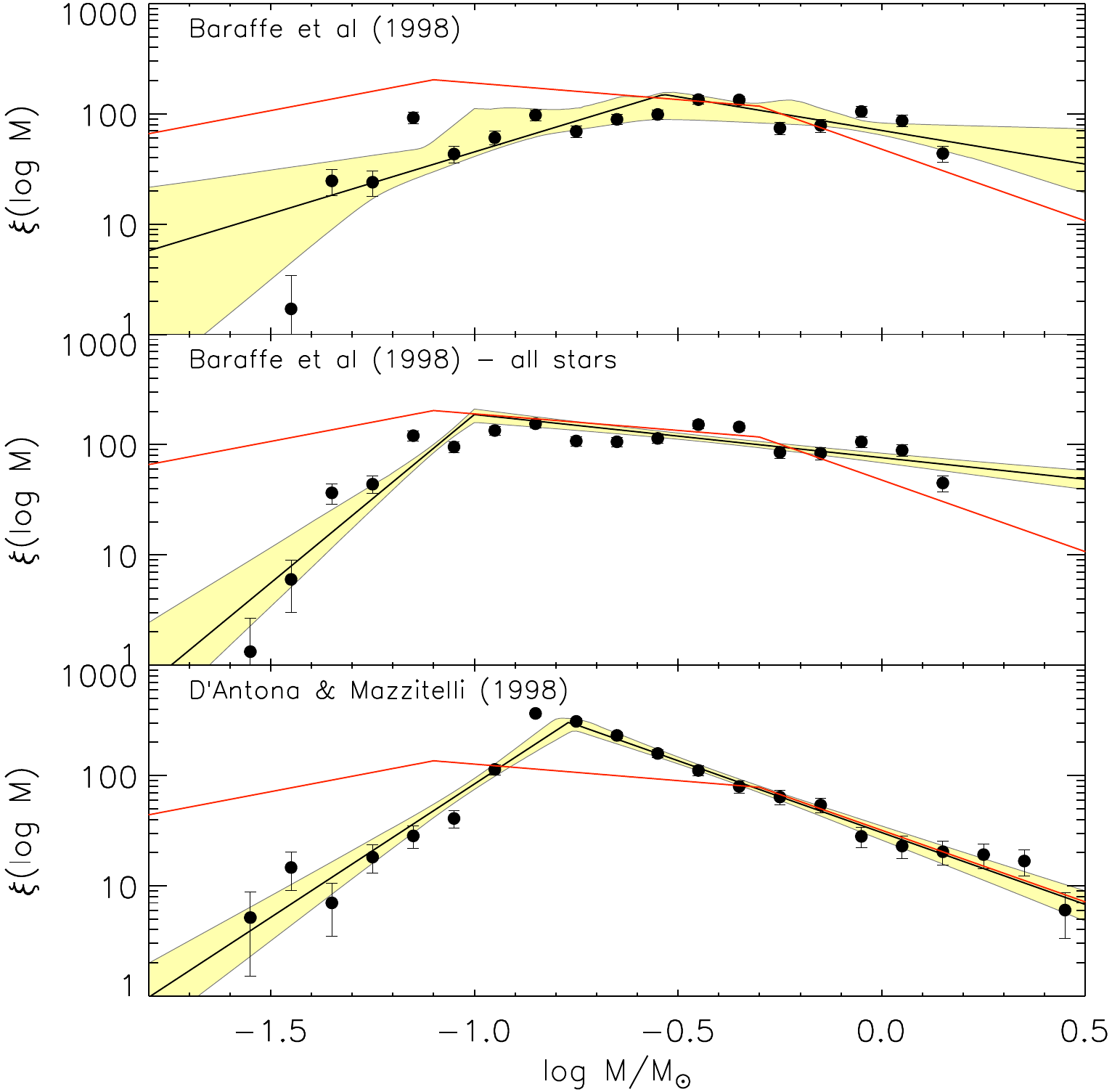}
\caption{The measured IMF for the ONC, fitted with a log-normal distribution (left panel) or a two-phase power law (right panel). The top and bottom panels represent the mass distributions obtained assuming \citet{bcah98} and \citet{dantona98} models respectively, whereas the center panels show the result assuming Baraffe models and including also stars located above the 1~Myr, whose mass has been extrapolated from their $\teff$ (see text). The shaded areas enclose the 90\% confidence interval for each fit. The red curves represent the IMF of \citet{chabrier2003} (left panel) and \citet{kroupa2001} (right panel); the red dashed line is the \citet{chabrier2003} system IMF.
\label{fig:IMF}}
\end{figure*}

\begin{deluxetable*}{|r|rr|rrr|}
\tablecaption{IMF}
\tablehead{\multicolumn{1}{c}{ } & \multicolumn{2}{c|}{lognormal} & \multicolumn{3}{c|}{2-phase powerlaw}  \\
\multicolumn{1}{|c|}{ } & \multicolumn{1}{c}{$\log m_c$} & \multicolumn{1}{c|}{$\sigma(\log m)$}  & \multicolumn{1}{c}{$\Gamma_1$} & \multicolumn{1}{c}{$\Gamma_2$} & \multicolumn{1}{c|}{$\log m_c$}
}
\startdata
BCAH98           & $-0.45\pm0.02$ & $0.44\pm0.05$ & $-1.12\pm0.90$ & $0.60\pm0.33$ & $-0.53\pm0.26$ \\
BCAH98 all stars & $-0.53\pm0.02$ & $0.39\pm0.03$ & $-3.05\pm0.53$ & $0.39\pm0.10$ & $-1.00\pm0.08$ \\
DM98             & $-0.56\pm0.02$ & $0.38\pm0.01$ & $-2.41\pm0.25$ & $1.30\pm0.09$ & $-0.77\pm0.02$
\enddata
\tablecomments{The power law exponents $\Gamma$ follow the standard for which the Salpeter slope is $\Gamma=1.35$}
\label{table:IMF}
\end{deluxetable*}

\begin{sidewaystable}
\centering
{\small
\begin{tabular}{|c|cc|ccc|cc|cc|}

\hline
\multicolumn{6}{|c}{} & \multicolumn{2}{|c|}{BCAH98} & \multicolumn{2}{|c|}{DM98} \\
\multicolumn{1}{|c|}{ID} & \multicolumn{1}{c}{R.A.} & \multicolumn{1}{c|}{Dec} & \multicolumn{1}{c}{$\log T_{\rm eff}$} & \multicolumn{1}{c}{$\log L_{\rm bol}$} & \multicolumn{1}{c|}{$A_V$} & \multicolumn{1}{c}{M} & \multicolumn{1}{c|}{$\log$ age} & \multicolumn{1}{c}{M} & \multicolumn{1}{c|}{$\log$ age} \\
\multicolumn{1}{|c|}{ } & \multicolumn{1}{c}{h m s} & \multicolumn{1}{c|}{$^\circ$ $\prime$ $\prime\prime$} & \multicolumn{1}{c}{(K)} & \multicolumn{1}{c}{($L_{\odot}$)} & \multicolumn{1}{c|}{mag} & \multicolumn{1}{c}{(M$_\odot$)} & \multicolumn{1|}{c|}{(yr)} & \multicolumn{1}{c}{(M$_\odot$)} & \multicolumn{1}{c|}{(yr)} \\

\hline
      1 &  05 34 15.10 &  -05 23 00.0    &    3.522 $\pm$   0.010    &   -1.553 $\pm$   0.025    &    0.000 $\pm$   0.307    &    0.243 $\pm$   0.053    &    7.305 $\pm$   0.172    &    0.255 $\pm$   0.038   &    7.308 $\pm$   0.111 \\
      2 &  05 34 17.26 &  -05 22 36.7    &    3.564 $\pm$   0.010    &   -1.565 $\pm$   0.046    &    0.770 $\pm$   0.238    &    0.481 $\pm$   0.028    &    8.235 $\pm$   0.175    &    0.447 $\pm$   0.034   &    8.196 $\pm$   0.297 \\
      3 &  05 34 17.29 & -05 22 48.0    &    3.492 $\pm$   0.001    &   -1.078 $\pm$   0.018    &    1.300 $\pm$   0.096    &    0.164 $\pm$   0.006    &    6.295 $\pm$   0.020    &    0.161 $\pm$   0.002   &    6.299 $\pm$   0.017 \\
      4 &  05 34 20.57 & -05 21 29.8    &    3.510 $\pm$   0.004    &   -2.312 $\pm$   0.036    &    0.971 $\pm$   0.187    &    0.168 $\pm$   0.014    &    8.090 $\pm$   0.105    &    0.185 $\pm$   0.014   &    8.260 $\pm$   0.216 \\
      5 &  05 34 22.40 & -05 22 26.9    &    3.477 $\pm$   0.002    &   -1.063 $\pm$   0.035    &    0.016 $\pm$   0.180            & $\sim$   0.098                        & $<$6    &    0.136 $\pm$   0.002   &    6.239 $\pm$   0.026 \\
      6 &  05 34 18.37 & -05 22 54.9    &    3.524 $\pm$   0.010    &   -2.068 $\pm$   0.043    &    0.051 $\pm$   0.375    &    0.230 $\pm$   0.053    &    8.030 $\pm$   0.264    &    0.244 $\pm$   0.041   &    8.135 $\pm$   0.294 \\
      7 &  05 34 20.78 & -05 23 29.1    &    3.530 $\pm$   0.002    &   -0.845 $\pm$   0.027    &    0.548 $\pm$   0.115    &    0.341 $\pm$   0.012    &    6.565 $\pm$   0.036    &    0.245 $\pm$   0.006   &    6.306 $\pm$   0.035 \\
      8 &  05 34 24.78 & -05 22 10.5    &    3.540 $\pm$   0.001    &   -0.643 $\pm$   0.021    &    0.266 $\pm$   0.086    &    0.429 $\pm$   0.011    &    6.470 $\pm$   0.024    &    0.275 $\pm$   0.006   &    6.142 $\pm$   0.025 \\
      9 &  05 34 25.64 & -05 21 57.6    &    3.484 $\pm$   0.001    &   -1.006 $\pm$   0.021    &    0.436 $\pm$   0.100    &    0.119 $\pm$   0.005    &    6.005 $\pm$   0.044    &    0.146 $\pm$   0.001   &    6.207 $\pm$   0.017 \\
     10 &  05 34 26.51 & -05 23 23.7    &    3.497 $\pm$   0.001    &   -0.502 $\pm$   0.013    &    0.836 $\pm$   0.055            & $\sim$   0.161                        & $<$6    &    0.147 $\pm$   0.001   &    5.182 $\pm$   0.027 \\
     ... &  ... & ...    &   ...  &  ...  &    ...        &...         & ...   &  ...  &    ... \\
\hline
\end{tabular} }

\caption{(This table is available in its entirety in a machine-readable form in the online journal. A portion is shown here for guidance regarding its form and content.) \label{table:parameters}}
\end{sidewaystable}

To derive the initial mass function (IMF) of the ONC from our observations,
we use pre-main-sequence evolutionary models to convert our $\teff$ and
$\log L$ from the H-R diagram into masses and ages. We do this using both
the \citet{dantona98} and \citet{bcah98} models
(Figure \ref{fig:HRD_with_models}).
In the case of Baraffe models, for a large fraction
(about 25\%) of sources we cannot assign masses and ages, as these stars
are located above the 1~Myr isochrone, the minimum age computed for this
family of models. This not only decreases our stellar sample, but also
biases our findings, in particular the mass distribution.
To overcome this selection effect,
we consider two cases for the mass estimates from Baraffe models: a) we
reject all source above the 1~Myr isochrone; b) we include them by
assigning a mass based on the $\teff$-mass relation of the 1~Myr
isochrone. Since for very-low mass stars the PMS evolutionary tracks are
nearly vertical in the H-R diagram, this approximation is fairly good. In
Table \ref{table:parameters} we present the derived stellar parameters
for the ONC sources, using both sets of models.

From these masses,
we derive the mass distribution $\xi(\log m)$ by binning the ONC members
in equally-spaced mass bins. For each mass bin, we account for its exact
completeness by adding the inverse of the completeness of
each source. We associate an uncertainty distribution to each measured
value of $\xi(\log m)$ equal to a Possion distribution of mean $\mu=N_i$,
where $N_i$ is the number of sources in the $i-$th bin, scaled by a factor
equal to the overall completeness correction for that bin. We stress that,
strictly speaking, our mass function is actually a ``system'' mass function
rather than a proper ``initial'' mass function, in the sense that we do
not account for unresolved binaries or multiple systems. This, however,
does not influence significantly our results, since the binary fraction
(accounting both bound systems and visual binaries) in Orion is small
($\lesssim 15\%$, \citealt{Padgett97,petr98,reipurth2007} ), and about half of
these are separated more than 1\arcsec, therefore resolved in
our observations.

It it well established \citep[e.g.,][]{bastian2010} that the IMF generally
follows a power-law in the intermediate- and high-mass range ($M \gtrsim
M_\odot$), whereas for low-mass stars and BDs---which is the region of
the mass spectrum most relevant for our study---this function can be
approximated with a shallower power law
$\xi(\log m)\propto m^{-(\Gamma+1)}$ \citep[e.g.,][]{kroupa2001} or
with a log-normal distribution $\xi(\log m) \propto e^{-(\log m - \log
m_{c})^{2}/2\sigma^{2}}$ \citep[e.g.,][]{chabrier2003}. We use both
forms to fit our measured IMF in the ONC.

We use a Monte Carlo simulation following \citet{da_rio2009LH95}
to account for the uncertainties in the measured star counts, as follows.
For every mass bin, we consider the error
bars with their statistical distribution, and generate a large number
($n=10,000$) of points drawn from the error distribution. Then an
unweighted fit is run on all these ($n$ times the number of bins) points. The
best-fit parameters are isolated using a Levenberg-Marquard minimization
algorithm. The uncertainty associated to the parameters has been computed
with a sampling technique as follows: for every one of the $m$ bins, we
randomly consider only one of the $n$ values previously simulated, i.e. a
random sample from the distribution describing the error bar of the bin,
and we fit the IMF function on these $m$ data points, deriving the best-fit
parameters. By iterating this selection and fit process 1000 times, we derive
1000 sets of parameters. The standard deviation of each parameter for the
1000 tests is the uncertainty in the estimate of the parameter itself.

The resulting fitted functions for both sets of evolutionary models are shown
in Figure \ref{fig:IMF}. In Table \ref{table:IMF} the fitting parameters are
reported: the characteristic mass and the width for the log-normal fit,
and the two power law slopes as well as the break point for the 2-phase
power law. We find that the two families of evolutionary models lead to
significant differences in the derived IMF. Whereas the Baraffe tracks
produce a smooth distribution, which appears well fitted by a log-normal
distribution characterized by a continuous change in the IMF slope, the
\citeauthor{dantona98} models lead to a more peaked distribution, with an
evident maximum at $M \sim 0.15$~M$_\odot$. The reason for this difference
can be attributed to the different shapes of the evolutionary tracks in
the very low mass range ($0.3$~M$_\odot\gtrsim M \gtrsim0.1$~M$_\odot$),
where the models of \citeauthor{dantona98} predict a larger area of the HRD
covered by adjacent tracks. The shape of the IMF derived
from the \citeauthor{bcah98} models does not depend significantly on
which method we adopt for inclusion of very young stars lying above
1~Myr isochrone (see above).

In Figure \ref{fig:IMF} we also compare our measured mass distribution
with standard reference IMFs. In particular we consider the disk IMF from
\citet{chabrier2003} which is expressed as a log-normal function, with a
characteristic mass of either 0.079~M$_\odot$ or 0.22~M$_\odot$ respectively
for the ``single objects'' IMF or the ``system'' IMF, the latter not
corrected for stellar multiplicity. Also, we consider the universal IMF of
\citet{kroupa2001}, which is a multi-phase power law with shallower slopes
in the VLMS and BD regime. Depending on the assumed
evolutionary models and the type of function fitted, our IMF peaks in the
range 0.1~M$_\odot$ $-$ 0.3~M$_\odot$, the higher values being larger than
that most often reported, even in the case of unresolved multiplicity. This
finding, as well as the high dependence on the evolutionary model family
used to derive stellar masses, is consistent with what we have found in our
previous work (Paper~I) limited to the stellar mass range.

A striking feature of our IMF is the steep decline in the BD mass range
(see Figure \ref{fig:IMF}); here our mass distribution is much lower
than the standard IMFs, implying a significant deficiency of substellar
objects in the ONC. We stress that this finding is not biased by detection
incompleteness, given that the derived mass distribution has been corrected
for completeness star by star, and the completeness function we have derived
in Section \ref{subsection:completenessHRD} remains fairly high ($\sim 50\%$)
even in the BD mass range. Moreover, even hypothesizing an overestimated
completeness for the smallest masses in our sample, the relative scarcity
of substellar members is evident up to about 0.1~M$_\odot$; there simply
are not many very cool, very low luminosity sources observed. Such a steep
decline of the mass function in the BD mass range differs also from
IMF determinations in other young regions. In fact the majority of young clusters have been reported to show substellar mass function compatible with the Kroupa or the Chabrier mass distributions. Some examples are Chamaeleon I \citet{luhman2007}, Upper Sco \citep{lodieu2007}, the $\sigma$ Ori region \citep{caballero2009}, $\lambda$ Ori \citep{bayo2011}, NGC~6611 \citep{oliveira2009}. Similar findings have also been reported for older open clusters \citep[e.g.,][]{bouvier2005}; in fact for the Praesepe cluster, \citet{Wang2011} measure even an increasing IMF down to $\sim 70$~M$_J$.

Previous studies of the IMF in the ONC also found a decrease in the
relative number of members in the substellar mass range, both in absolute
number and relative to the standard IMFs. Studies based on photometry alone
\citep[e.g.,][]{hillenbrand-carpenter2000,muench2002} using the
conversion of near infrared LFs, find a measured IMF slope below the H-burning
limit not steeper than $\Gamma\sim-1$, which is the shallowest value
we derive assuming \citet{bcah98} models, and up to 2 units flatter than
what we derive with \citet{dantona98} tracks. Also, \citet{slesnick04}
finds an IMF for the ONC rapidly decreasing below the H-burning limit,
but then flattening in the BD mass range.

\citet{muench2002} reports a secondary peak in the mass distribution at
about the deuterium burning limit ($\sim 0.17$ M$_\odot$); \citet{slesnick04}
tentatively find a similar feature, although with less significance and
at a slightly higher mass. This mass roughly corresponds to our detection
limit. However, from our results, we do not find any evidence of either a
flattening of the IMF or an increasing number of sources for the lowest
mass bins. We interpret this inconsistency as an inaccurate estimate of
the background contamination in previous works, which we believe is much
improved through the methodology that we have developed here.

Based on our measured IMF, complemented in the intermediate- and high
mass range using the results from \citet{hillenbrand97}, we assess a total
stellar mass for the ONC of about $10^3$~M$_\odot$.

\acknowledgements
The authors acknowledge the Max-Planck Society (MPG) and the Max-Planck Institute for Astronomy (Heidelberg, Germany) for the telescope time.

The IMF research of Thomas Henning is supported by Sonderforschungsbereich SFB 881
"The Milky Way System" (subproject B6) of the German Research
Foundation (DFG).

{\it Facilities:} \facility{Max Planck:2.2m, HST}


\begin{thebibliography}{}
\bibitem[Allard et al.(2010)]{allard2010} Allard, F., Homeier, D., \& Freytag, B.\ 2010, ASP Conf. Ser., Cool Star 16, {\em in press} (arXiv:1011.5405)
\bibitem[Allard et al.(2000)]{allartio} Allard, F., Hauschildt, P.~H., \& Schwenke, D.\ 2000, \apj, 540, 1005
\bibitem[Baraffe et al.(1998)]{bcah98} Baraffe, I., Chabrier, G., Allard, F., et al.\ 1998, \aap, 337, 403
\bibitem[Bastian et al.(2010)]{bastian2010} Bastian, N., Covey, K.~R., \& Meyer, M.~R.\ 2010, \araa, 48, 339
\bibitem[Bayo et al.(2011)]{bayo2011} Bayo, A., Barrado, D., Stauffer, J., et al.\ 2011, arXiv:1109.4917
\bibitem[Bergin et al.(1996)]{Bergin1996} Bergin, E.~A., Snell, R.~L., \& Goldsmith, P.~F.\ 1996, \apj, 460, 343
\bibitem[Bessell et al.(1998)]{bessel98} Bessell, M.~S., Castelli, F., \& Plez, B.\ 1998, \aap, 333, 231
\bibitem[Bouvier et al.(2005)]{bouvier2005} Bouvier, J., Moraux, E., \& Stauffer, J.\ 2005, The Initial Mass Function 50 Years Later, 327, 61
\bibitem[Caballero(2009)]{caballero2009} Caballero, J.~A.\ 2009, American Institute of Physics Conference Series, 1094, 912
\bibitem[Calvet \& Gullbring(1998)]{calvetgullbring98} Calvet, N., \& Gullbring, E.\ 1998, \apj, 509, 802
\bibitem[Cardelli et al.(1989)]{cardelli} Cardelli, J.~A., Clayton, G.~C., \& Mathis, J.~S.\ 1989, \apj, 345, 245
\bibitem[Chabrier(2003)]{chabrier2003} Chabrier, G.\ 2003, \pasp, 115, 763
\bibitem[Cohen \& Kuhi(1979)]{cohencuhi79} Cohen, M., \& Kuhi, L.~V.\ 1979, \apjs, 41, 743
\bibitem[Costero \& Peimbert(1970)]{costero70} Costero, R., \& Peimbert, M.\ 1970, Boletin de los Observatorios Tonantzintla y Tacubaya, 5, 229
\bibitem[D'Antona \& Mazzitelli(1998)]{dantona98} D'Antona, F., \& Mazzitelli, I.\ 1998, Brown Dwarfs and Extrasolar Planets, 134, 442
\bibitem[Da Rio et al.(2010)]{paperII} Da Rio, N., Robberto, M., Soderblom, D.~R., et al.\ 2010, \apj, 722, 1092
\bibitem[Da Rio et al.(2009b)]{paperI} Da Rio, N., Robberto, M., Soderblom, D.~R., et al.\ 2009, \apjs, 183, 261
\bibitem[Da Rio et al.(2009a)]{da_rio2009LH95} Da Rio, N., Gouliermis, D.~A., \& Henning, T.\ 2009, \apj, 696, 528
\bibitem[De Marchi et al. (2011)]{demarchi2011} De Marchi, G., Panagia, N., Guarcello, et al.\ 2010, \aap \emph{submitted}
\bibitem[Fischer et al.(2011)]{fischer2011} Fischer, W., Edwards, S., Hillenbrand, L., et al.\ 2011, \apj, 730, 73
\bibitem[Flower(1996)]{flower96} Flower, P.~J.\ 1996, \apj, 469, 355
\bibitem[Guarcello et al.(2010)]{guarcello2010} Guarcello, M.~G., Damiani, F., Micela, G., et al.\ 2010, \aap, 521, A18
\bibitem[Hillenbrand(1997)]{hillenbrand97} Hillenbrand, L.~A.\ 1997, \aj, 113, 1733
\bibitem[Hillenbrand \& Carpenter(2000)]{hillenbrand-carpenter2000} Hillenbrand, L.~A., \& Carpenter, J.~M.\ 2000, \apj, 540, 236
\bibitem[Kraus \& Hillenbrand(2009)]{kraus-hillenbrand2009} Kraus, A.~L., \& Hillenbrand, L.~A.\ 2009, \apj, 703, 1511
\bibitem[Kroupa(2002)]{kroupascience} Kroupa, P.\ 2002, Science, 295, 82
\bibitem[Kroupa(2001)]{kroupa2001} Kroupa, P.\ 2001, \mnras, 322, 231
\bibitem[Lodieu et al.(2007)]{lodieu2007} Lodieu, N., Hambly, N.~C., Jameson, R.~F., et al.\ 2007, \mnras, 374, 372
\bibitem[Lucas et al.(2006)]{lucas2006} Lucas, P.~W., Weights, D.~J., Roche, P.~F., et al.\ 2006, \mnras, 373, L60
\bibitem[Lucas et al.(2005)]{lucas2005} Lucas, P.~W., Roche, P.~F., \& Tamura, M.\ 2005, \mnras, 361, 211
\bibitem[Lucas et al.(2001)]{lucas2001} Lucas, P.~W., Roche, P.~F., Allard, F., et al.\ 2001, \mnras, 326, 695
\bibitem[Lucas \& Roche(2000)]{lucasroche2000} Lucas, P.~W., \& Roche, P.~F.\ 2000, \mnras, 314, 858
\bibitem[Luhman et al.(2000)]{luhman2000} Luhman, K.~L., Rieke, G.~H., Young, E.~T., et al.\ 2000, \apj, 540, 1016
\bibitem[Luhman(2007)]{luhman2007} Luhman, K.~L.\ 2007, \apjs, 173, 104
\bibitem[Menten et al.(2007)]{menten2007} Menten, K.~M., Reid, M.~J., Forbrich, J., et al.\ 2007, \aap, 474, 515
\bibitem[Meyer et al.(1997)]{meyer97} Meyer, M.~R., Calvet, N., \& Hillenbrand, L.~A.\ 1997, \aj, 114, 288
\bibitem[Muench et al.(2002)]{muench2002} Muench, A.~A., Lada, E.~A., Lada, C.~J., et al.\ 2002, \apj, 573, 366
\bibitem[Muench et al.(2000)]{muench2000} Muench, A.~A., Lada, E.~A., \& Lada, C.~J.\ 2000, \apj, 533, 358
\bibitem[Oliveira et al.(2009)]{oliveira2009} Oliveira, J.~M., Jeffries, R.~D., \& van Loon, J.~T.\ 2009, \mnras, 392, 1034
\bibitem[Padgett et al.(1997)]{Padgett97} Padgett, D.~L., Strom, S.~E., \& Ghez, A.\ 1997, \apj, 477, 705
\bibitem[Palla \& Stahler(1999)]{pallastahler99} Palla, F., \& Stahler, S.~W.\ 1999, \apj, 525, 772
\bibitem[Petr et al.(1998)]{petr98} Petr, M.~G., Coud{\'e} du Foresto, V., Beckwith, S.~V.~W., et al.\ 1998, \apj, 500, 825
\bibitem[Reipurth et al.(2007)]{reipurth2007} Reipurth, B., Guimar{\~a}es, M.~M., Connelley, M.~S., et al. 2007, \aj, 134, 2272
\bibitem[Reggiani et al.(2011)]{reggiani2011} Reggiani, M., Robberto, M., Da Rio, N., et al.\ 2011, \aap, 534, A83
\bibitem[Riddick et al.(2007)]{riddick2007} Riddick, F.~C., Roche, P.~F., \& Lucas, P.~W.\ 2007, \mnras, 381, 1077
\bibitem[Robberto et al.(2010)]{robberto2010} Robberto, M., Soderblom, D.~R., Scandariato G., et al.\ 2010, \aj, 139, 950
\bibitem[Robberto et al.(2005)]{robberto2005treasury} Robberto, M., et al.\ 2005, Bulletin of the American Astronomical Society, 37, \#146.01
\bibitem[Salpeter(1955)]{salpeter55} Salpeter, E.~E.\ 1955, \apj, 121, 161
\bibitem[Scandariato et al.(2011)]{scandariato_extinction} Scandariato, G., Robberto, M., Pagano, I., et al.\ 2011, \aap, 533, A38
\bibitem[Siess et al.(2000)]{siess2000} Siess, L., Dufour, E., Forestini, M.\ 2000, \aap, 358, 593
\bibitem[Slesnick et al.(2004)]{slesnick04} Slesnick, C.~L., Hillenbrand, L.~A., \& Carpenter, J.~M.\ 2004, \apj, 610, 1045
\bibitem[Stetson(1987)]{stetson87} Stetson, P.~B.\ 1987, \pasp, 99, 191
\bibitem[Vandame (2004)]{vandame04} Vandame, B., PhD thesis, Nice University, 2004
\bibitem[Wang et al.(2011)]{Wang2011} Wang, W., Boudreault, S., Goldman, B., et al.\ 2011, \aap, 531, A164
\bibitem[Weights et al.(2009)]{weights2009} Weights, D.~J., Lucas, P.~W., Roche, P.~F., et al.\ 2009, \mnras, 392, 817


\end{thebibliography}
\end{document}